\documentclass[prd,superscriptaddress,twocolumn,preprintnumbers,showpacs,nofootinbib]{revtex4}
\usepackage{amssymb,amsmath,graphicx,epsfig,subfig,color} 

\def\beq {\begin{equation}}
\def\eeq {\end{equation}}
\def\bea {\begin{eqnarray}}
\def\eea {\end{eqnarray}}

\begin{document}
\title{Large mass splittings for fourth generation fermions allowed by LHC Higgs boson exclusion} 

\author{Amol Dighe}
\email{amol@theory.tifr.res.in} 
\affiliation{Tata Institute of 
Fundamental Research, 1, Homi Bhabha Road, Mumbai 400 005, India}

\author{Diptimoy Ghosh}
\email{diptimoyghosh@theory.tifr.res.in}
\affiliation{Tata Institute of 
Fundamental Research, 1, Homi Bhabha Road, Mumbai 400 005, India}

\author{Rohini M Godbole}
\email{rohini@cts.iisc.ernet.in}
\affiliation{Centre for High Energy Physics, Indian Institute of 
Science, Bangalore 560 012, India}

\author{Arun Prasath}
\email{arunprasath@cts.iisc.ernet.in}
\affiliation{Centre for High Energy Physics, Indian Institute of 
Science, Bangalore 560 012, India}
\date{\today}
\begin{abstract} 
In the context of the standard model with a fourth generation, we explore the 
allowed mass spectra in the fourth-generation quark and lepton sectors as 
functions of the Higgs mass. Using the constraints from unitarity and oblique parameters, we show that a heavy Higgs allows large mass splittings in these sectors, opening up new decay 
channels involving $W$ emission. Assuming that the hints for a light Higgs do not yet constitute an evidence, we work in a scenario where a heavy Higgs is viable. A Higgs heavier than $\sim 800$ GeV would in fact necessitate either a heavy quark decay channel $t'\rightarrow b'W/b'\rightarrow t' W$ or a heavy lepton decay channel $\tau '\rightarrow \nu ' W$ as long as the mixing between the third and fourth generations is small. This mixing tends to suppress the mass splittings and hence the $W$-emission channels. The possibility of the $W$-emission channel could substantially change the search strategies of fourth-generation fermions at the LHC and impact the currently reported mass limits.
\end{abstract}

\preprint{TIFR/TH/12-11}
\maketitle
\section{Introduction}

The number of fermion generations in the standard model (SM) is three,
though there is no fundamental principle restricting it to this number. The data on $Z$ decay only puts a lower bound of $m_Z/2$ on their 
masses \cite{Nakamura:2010zzi}. Direct searches at the Tevatron 
\cite{Nakamura:2010zzi,Lister:2008is,PhysRevLett.104.091801,PhysRevLett.106.141803,PhysRevLett.107.082001,Heintz:2012kt}
and the LHC \cite{Aad:2011vj,Bousson:2012sp,Aad:2012us,Aad:2012bb,Chatrchyan2011204,Dahmes:2012tb,cms:2012,Rahatlou:2012hm} further put lower bounds on the masses of fourth-generation charged fermions, by 
virtue of them not having been observed at these colliders. These 
limits are subject to certain assumptions about the decay channels for 
these quarks. The most conservative, almost model-independent limits for 
fourth-generation quarks are given in \cite{Flacco:2011ym}. 
Indirect limits on the masses and mixing of these 
fermions are also obtained through the measurements of the oblique 
parameters $S, T, U$ \cite{Baak:2011ze}. Theoretical constraints like 
the perturbativity of the Yukawa couplings and the perturbative unitarity 
of heavy fermion scattering amplitudes \cite{Chanowitz:1978uj} bound the 
masses from above. In spite of the rather strong bounds from all the above 
directions, there is still parameter space available for the fourth generation
that is consistent with all the data \cite{
Evans:1994an,Novikov:1994zg,He:2001tp,Novikov:2001md,Novikov:2002tk,PhysRevD.68.093012,Hung:2007ak,Chanowitz:2009mz,Novikov:2009kc,Erler:2010sk,Eberhardt:2010bm,Eberhardt:2012sb}. The fourth-generation 
scenario (SM4) is thus still viable even after the recent Tevatron and 
LHC results \cite{Rozanov:2010xi,Cetin:2011aa}.

The discovery of a fourth-generation of fermions will have profound 
phenomenological consequences \cite{Holdom:2009rf}. 
Some of the experimental observations that deviate somewhat from the SM 
expectations -- like the CP-violating phases in the neutral $B$ mixing
\cite{Hou:2005hd,Hou:2006mx,Soni:2008bc,Soni:2010xh} -- could be 
interpreted as radiative effects by the fourth-generation fermions. 
The implications of a fourth family for observables in charmed decays 
\cite{Buras:2010nd} and lepton-flavor violating decays \cite{Buras:2010cp} as well as flavor constraints on the quark sector\cite{Bobrowski:2009ng} have also been discussed.

The Cabibbo-Kobayashi-Maskawa (CKM) structure of 4 generations, with 3 observable phases and large 
Yukawa couplings, may also provide enough source of CP violation for the 
matter-antimatter asymmetry of the Universe, although the issue of the 
order of the electroweak phase transition still has to be resolved 
\cite{Hou:2008xd}. 
The existence of a fourth-generation is also intimately connected with 
the Higgs physics. 
The Higgs in the SM4 with a mass of about 800 GeV  
is consistent with precision electroweak (EW)data when 
$m_{t'},m_{b'}\sim 500$ GeV, 
with the mixing between third and fourth-generation quarks of 
the order of 0.1 \cite{Chanowitz:2009mz}. 
This raises an interesting possibility  
of Higgs being a composite scalar of fourth-generation quarks 
\cite{Hung:2009hy,BarShalom:2010bh,Hung:2010xh} with interesting phenomenological 
implications including an enhancement of flavor-changing as well as 
flavor-diagonal Higgs decays into third and fourth-generation fermions. 
Implications of a strongly interacting fourth-generation quark sector on 
LHC Higgs searches has been discussed in \cite{B2012381}. Phenomenology of 
the lepton sector of the SM4 has been studied in \cite{Carpenter:2010sm,
Carpenter:2010bs,Carpenter:2010dt,Schmidt:2011jp}.

The Higgs production cross section at hadron colliders is affected 
strongly by the fourth-generation through the $gg \to h$ channel 
due to the heavy masses of the new fermions 
\cite{Djouadi:2005gi,Anastasiou:2011qw,Passarino:2011kv,Denner:2011vt}.
The branching ratios of Higgs into different channels are affected too   
\cite{Djouadi:1997rj,Djouadi:2005gi,Denner:2011vt}.
As a consequence, the direct search limits on the Higgs mass are stronger 
in the presence of the fourth-generation. 
Higgs production and decay cross sections in the context of four generations,
with next-to-leading order EW and QCD corrections, have been
calculated in \cite{Anastasiou:2011qw,Passarino:2011kv,Denner:2011vt}. 
The production cross section is enhanced for a light Higgs 
($m_h<200$ GeV) by an order of magnitude. However the enhancement may be somewhat
reduced for a heavier Higgs \cite{2010EPJC...66..119B}. 
The ATLAS experiment at the LHC 
has excluded the Higgs boson of SM4 with a mass between 119 GeV and 593 GeV 
\cite{ATLAS-CONF-2011-135}, while the CMS exclusion limits are from 120 GeV 
to 600 GeV \cite{CMS:2011}. These experiments include the one loop EW 
corrections to Higgs production from the fourth-generation fermions\footnote{These bounds can be circumvented either by a suitable extension of the 
scalar sector \cite{He:2011ti} or by having Higgs decay to stable invisible 
particles which could be candidates for dark matter\cite{PhysRevD.68.054027,Melfo:2011ie,Carpenter:2011wb,Cetin:2011fp,Borah:2011ve}. However this is 
not the minimal SM4 we focus on in this paper.}. 

Recently, experiments at the Tevatron and the LHC have reported an excess of events around $m_h\approx 125 \ \mathrm{GeV}$ with a local  significance of about $2\sigma - 3\sigma$ in the search for the standard model Higgs boson \cite{TEVNPH:2012ab,ATLAS:2012ae,Chatrchyan:2012tx}. Reference\cite{Djouadi:2012ae} interprets these ``hints'' for a light Higgs in the framework of SM4. We on the otherhand, assume that these hints do not yet constitute an evidence and \textit{may} well be a statistical fluctuation. Under this assumption, a heavy Higgs with mass $\gtrsim 600 \ \mathrm{GeV}$ is a viable scenario.

Recent explorations of possible effects of a fourth-generation on the 
Higgs mass, precision observables, quark mixing matrix, and flavor-changing 
neutral current phenomena have yielded interesting results. 
It has been pointed out \cite{Kribs:2007nz} that the existence of a fourth 
generation allows for a heavier Higgs to be consistent with the 
precision measurements. The constraints on the mixing between the third and 
fourth generation have been obtained from the precision EW data 
\cite{Chanowitz:2009mz}, and a fit to the flavor-physics data 
\cite{Alok:2010zj}. 
The latter shows that, while the mixing of the fourth-generation quarks to 
the three SM generations is consistent with zero and restricted to be small, 
observable effects on $K$ and $B$ decays are still possible. Large masses of the 
fourth-generation fermions lead to nonperturbativity of Yukawa couplings at a 
low scale $\Lambda \ll M_{GUT}$ as well as instability of the vacuum. 
This has been investigated in the context of models without supersymmetry (SUSY) 
\cite{Kribs:2007nz,Hung:2009hy,Hung:2009ia,Knochel:2011ng,Ishiwata:2011hr,Wingerter:2011dk} 
and with SUSY \cite{Godbole:2009sy}, after taking into account various bounds 
from precision EW data as well as collider and direct search experiments on 
sequential heavy fermions.

In this article, we revisit the electroweak precision constraints from the
oblique parameters $S, T, U$ on the fourth-generation, taking into account 
the mixing with the third generation. We perform a $\chi^2$ analysis, varying 
the fourth-generation quark as well as lepton masses in their experimentally 
allowed ranges, and obtain a quantitative measure for the fourth-generation 
fermion mass spectrum preferred by the measurements of these parameters. 
In the light of the heavy Higgs preferred by the LHC data, we focus on the 
implications of a heavy Higgs for the mass spectrum. We also study the effect 
of the mixing between third and fourth-generation on this mass spectrum,
and try to understand these effects analytically. 
As we will see later, the correlation between the mass splittings in quark 
and lepton sectors is strongly influenced by this mixing angle.

The paper is organized as follows. In Sec.~\ref{sec:constraints}, we discuss 
the bounds on the fourth-generation fermion masses from direct searches and 
the theoretical requirement of perturbative unitarity. We also analyze the 
structure of constraints from the measurements of oblique parameters. 
In Sec.~\ref{sec:numerical}, we perform a $\chi^2$-fit to the oblique parameters 
and obtain constraints on the mass splittings in the quark and lepton sectors, 
focusing on a heavy Higgs. Sec.~\ref{sec:collider} discusses the collider 
implications, while Sec.~\ref{sec:concl} summarizes our results.

\section{Constraints on the fourth-generation fermion masses}
\label{sec:constraints}

\subsection{Lower bounds on masses from direct searches}

The direct search constraints presented by CDF \cite{Lister:2008is,
PhysRevLett.104.091801} on the masses of $t'$ and $b'$ quarks have been generalized 
to more general cases of quark mixing by \cite{Flacco:2011ym}, and a lower bound  
$m_{t'}$,$m_{b'}>290$ GeV has been obtained. The currently quoted exclusion bounds by CDF, D\O{}, CMS and ATLAS collaborations \cite{Heintz:2012kt,Bousson:2012sp,Aad:2012us,Aad:2012bb,Dahmes:2012tb,Rahatlou:2012hm} are 400-500 GeV, however as stated earlier, they are based on specific 
assumptions on branching ratios of the fourth-generation quarks and mass differences 
between the fourth-generation fermions.

The limits on the masses of heavy charged fermions are obtained from the nonobservation of their expected decay modes. The choice of analyzed decay modes affects the bounds to a large extent. Since we would like our results to be 
independent of assumptions about the mixing angles, mass differences, and hence branching ratios, in our 
analysis we shall use the bounds from \cite{Flacco:2011ym} for the quark masses. For the fourth-generation leptons $\tau'$ and $\nu'$, we take 
the bounds $m_{\tau'}>101.0 $ GeV and $m_{\nu _4}>45.0$ GeV \cite{Nakamura:2010zzi}. The mixing of the fourth-generation leptons is restricted to be very small, so it would not affect our analysis.

\subsection{Upper bounds on masses from unitarity}

The direct search constraints imply that the fourth-generation quarks are 
necessarily heavy ($m_F \gg M_W,M_Z$). For such heavy fermions $F$, the tree-level 
amplitudes of certain processes like $F \bar{F} \rightarrow F \bar{F}, WW, ZZ, ZH, HH$, 
in a spontaneously broken $SU(2)_L\times U(1)_Y$ gauge theory, tend to a constant 
value $G_F m^{2}_{F}$ at center-of-mass energies $s \gg m_{F}^{2}$. For large value of 
$m_F$, the term  $G_F m^{2}_{F}$ can be ${\cal O}(1)$. In that case, the tree-level 
unitarity of the S-matrix is saturated and in order to regain a unitary S-matrix, higher 
order amplitudes need to contribute significantly. This necessitates a strong coupling 
of these fermions to the gauge bosons, which makes the perturbation theory unreliable. 
This was first studied in the context of the SM with ultraheavy fermions in \cite{Chanowitz:1978uj,Chanowitz:1978mv}.The corresponding analysis in the 
context of the minimal supersymmetric standard model with a sequential fourth 
generation  was performed in \cite{Dawson:2010jx}, in the limit of vanishing mixing between the fourth 
generation quarks and the first three generations.

We reevaluate the bounds given in \cite{Chanowitz:1978mv}, 
considering the $J=0$ partial-wave channel
of the tree-level amplitudes of the color-neutral and charge-neutral processes 
$F_i \bar{F}_i\rightarrow F_j \bar{F}_j$.
In \cite{Chanowitz:1978mv} only the amplitudes involving two heavy fermions of a $SU(2)_L$ doublet were analyzed. The second $SU(2)_L$ doublet of heavy fermions only provided a source for mixing included in  the analysis. However,
we include all the relevant channels involving all the heavy quarks -$t'$, $b'$ and $t$, 
and take into account mixing between the third and fourth-generations. 
The lowest critical value of the fermion 
mass is obtained by equating the largest eigenvalue of this submatrix to unity. Expressions 
for the partial-wave matrices are given in the Appendix. The results  are 
shown in Fig. \ref{unit-comb} for $\sin \theta _{34}=0.0, 0.3$. One can easily observe 
from the figure an improvement of about 6\% in the bounds compared to those in 
\cite{Chanowitz:1978mv}. The bounds are affected due to the inclusion of the top quark in the analysis, which introduces more scattering channels. It may be seen from Fig. \ref{unit-comb} that the bounds are not very sensitive to the actual value of the mixing.

In the lepton sector, only fourth-generation leptons ($\tau'$ and $\nu_{\tau'}$) are 
relevant for the perturbative unitarity constraints as all the first three generation 
fermions are light compared to $M_W$. The mixing between the fourth-generation leptons 
and the first three generations is constrained by experimental bounds on lepton- flavor 
violating processes \cite{Buras:2010cp,Lacker:2010zz,Deshpande:2011uv}. The 2$\sigma$ 
lower bound on the (4,4) element of the Pontecarvo-Maki-Nakagawa-Sakata matrix in SM4, $U_{\tau'\nu'}$, is very 
close to unity: $|U_{\tau'\nu'}| > 0.9934$ \cite{Lacker:2010zz}. Moreover, we do not 
have any heavy fermion in the first three generations in contrast to the quark sector 
which has the top quark. Therefore, we do not have any new channel in addition to those 
that are considered in \cite{Chanowitz:1978mv}.Hence we do not expect any improvement over 
the bounds given in \cite{Chanowitz:1978mv} in the case of leptons. 

\begin{figure}[]
\begin{center}
\includegraphics[scale=0.4]{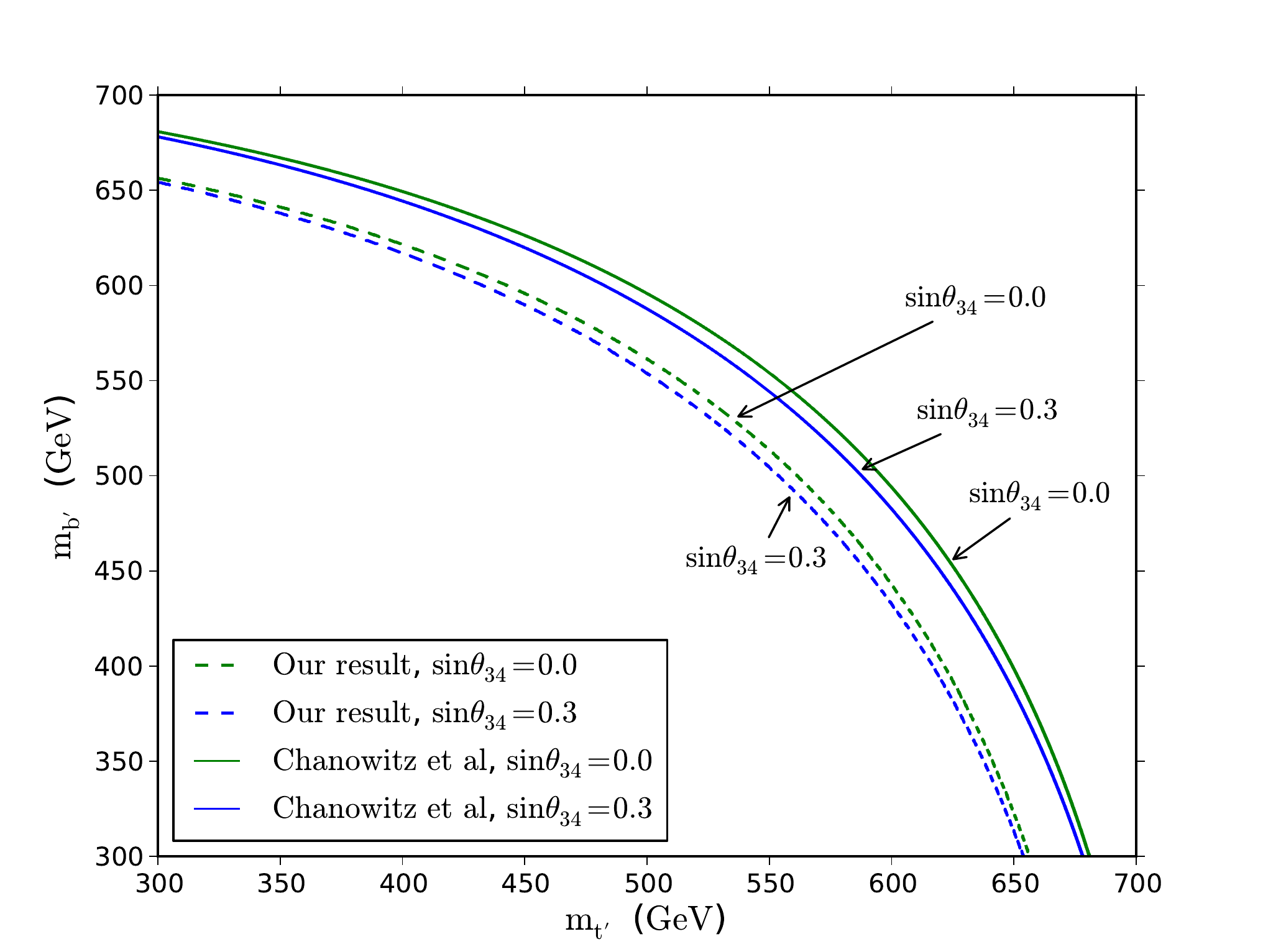}
\end{center}
\caption{The perturbative unitarity bounds on ($m_{t'},m_{b'}$) are shown above as 
dashed-lines for $\sin \theta _{34}=0.0$ (green/light) and 0.3 (blue/dark). The 
solid lines are obtained from the analytical expression for the unitarity bound 
given in \cite{Chanowitz:1978mv} by substituting the respective values of 
$\sin \theta _{34}$. The corresponding bounds on the fourth-generation lepton masses 
are $m_{\tau^{\prime},\,\nu^{\prime}} < 1.2 $ TeV \cite{Chanowitz:1978mv}.}
\label{unit-comb}
\end{figure}
%
%
%
\subsection{Constraints from oblique parameters}

The oblique parameters $S, T, U$ \cite{Peskin:1991sw} are sensitive probes of the 
fourth-generation masses and mixing pattern. The definitions of $S, T, U$ parameters 
are given by 
\begin{eqnarray}
S & = & 16\pi[\Pi'_{33}(0)-\Pi'_{3Q}(0)] \, , \\
T & = & \frac{4\pi}{s^2_W(1-s^2_W)m^2_Z}[\Pi _{11}(0)-\Pi_{33}(0)] \, , \\
U & = & 16\pi[\Pi'_{11}(0)-\Pi'_{33}(0)] \, ,
\end{eqnarray}
where $s^2_W=\sin ^2\theta _W $, and $\Pi(q^2)$ are the vacuum polarization
$\Pi$-functions. The suffixes 1, 2, 3 refer to the generators of $SU(2)_L$, 
and the suffix $Q$ to that of the electromagnetic current. The contribution to these 
parameters from a fermion doublet $(u, d)$ is \cite{Peskin:1991sw} 
\begin{eqnarray}
S(x_{u},x_{d}) & = & \frac{N_C}{6\pi}\left[1-Y\log 
\left( \frac{x_{u}}{x_{d}}\right)\right] \, , \hspace{3cm} \\
T(x_{u},x_{d}) & = & \frac{N_C}{16\pi s^2_W(1-s^2_W)} \times \nonumber 
\label{S-xuxd} \\
& & \left[x_{u}+x_{d}-\frac{2x_{u}x_{d}}{x_{u}-x_{d}}
\log\left(\frac{x_{u}}{x_{d}}\right)\right] \, , 
\label{T-x1x2} \\
U(x_{u},x_{d}) & = & 
\frac{N_C}{6\pi}\left[-\frac{5x_{u}^2-22x_{u}x_{d}+5x_{d}^2}{3(x_{u}-x_{d})^2}+
\right. \nonumber \\
& &\frac{x_{u}^3-3x_{u}^2x_{d}-3x_{u}x_{d}^2+x_{d}^3}{(x_{u}-x_{d})^3}
\left.\log\left(\frac{x_{u}}{x_{d}}\right)\right] \, ,
\end{eqnarray}
where $x_f \equiv m^2_f/m^2_Z$. Here $N_C$ is the number of colors of the fermions 
($N_C =3$ for quarks and $N_C=1$ for leptons), and Y is the hypercharge of the fermion 
doublet. Note that when the quarks in the doublet are almost degenerate, i.e. 
$\Delta \equiv |m_u - m_d| \ll m_u, m_d$, 
\begin{equation}
T(x_u,x_d)=\frac{1}{12\pi s^2_W(1-s^2_W)}\left(\frac{\Delta ^2}{m^2_Z}\right)\; .
\label{Tdeg}
\end{equation}
These expressions can be readily generalized to the case of additional sequential 
generations.

Following Gfitter \cite{Baak:2011ze}, we fix the masses of the top quark and the 
Higgs boson at their reference values of $\tilde{m}_t = 173.1$ GeV and 
$\tilde{m}_{h} = 120$ GeV. With these masses, in the SM we have 
$S =0$, $T = 0$ and $U = 0$. The deviation from these values are denoted by $\Delta S$, 
$\Delta T$, $\Delta U$ respectively. The effect of the Higgs mass appears through the 
dependence 
\cite{Peskin:1991sw}
\begin{eqnarray}
\Delta S_H & = & \frac{1}{6\pi}\log\left(\frac{m_h}{\tilde{m}_{h}}\right) \, , 
\label{S-higgs}\\
\Delta T_H & = & -\frac{3}{8\pi(1-s^2_W)}
\log\left(\frac{m_h}{\tilde{m}_{h}}\right) \, ,
\label{T:higgs} \\
\Delta U_H & \approx & 0 \; .
\end{eqnarray}

The contribution from the fourth-generation of fermions to these parameters, 
after taking into account the mixing between third and fourth-generation 
of quarks, can be expressed as 
\begin{eqnarray}
\Delta S_4 & = & S(x_{t'},x_{b'})+S(x_{\nu'},x_{\tau'}) \, , \\
\Delta T_4 & = & -s^2_{34}T(x_t,x_b)+s^2_{34}T(x_{t'},x_b)+
s^2_{34}T(x_t,x_{b'})+ \nonumber \\
&& c^2_{34}T(x_{t'},x_{b'})+T(x_{\nu'},x_{\tau'}) \, , 
\label{T:fermion} \\
\Delta U_4 & = & -s^2_{34}U(x_t,x_b)+s^2_{34}U(x_{t'},x_b)+s^2_{34}U(x_t,x_{b'})
+ \nonumber \\
&&c^2_{34}U(x_{t'},x_{b'})+U(x_{\nu'},x_{\tau'}) \, ,
\end{eqnarray}
where $s^2_{34}=\sin ^2\theta _{34}$, $c^2_{34}=\cos ^2\theta _{34}$ and 
$\theta _{34}$ is the mixing angle between the third and the fourth-generation 
quarks. Note that here we neglect the mixing of the fourth-generation quarks 
with the first two generations, since the bounds on this mixing are rather strong 
\cite{Alok:2010zj}. We also neglect any mixing of the fourth leptonic generation. 
We work in an approximation in which we neglect the nonoblique corrections to precision 
EW observables. This allows us to use the $S, T, U$ values provided by fits to the 
precision EW observables, for example the ones provided by the Gfitter group 
\cite{Baak:2011ze}.

When the mixing of the fourth and third generation quarks is nonzero, the decay width 
for $Z\rightarrow b \bar{b} $ receives contribution from fourth-generation quarks through 
vertex corrections,  in addition to the oblique corrections. To study the effect of fourth 
generation fermions on the precision EW observables in general mixing scenarios, one should, in principle
include both the vertex and the oblique corrections to precision EW observables 
\cite{Gonzalez:2011he}. However, as mentioned above , in order to use the Gfitter results on $S,T,U$ which were obtained in the limit of vanishing mixing ,we use only the values of the mixing angles that are consistent with 
the $Z\rightarrow b \bar{b}$ constraints \cite{Alok:2010zj}.

In our analysis, we evaluate the $S, T, U$ parameters numerically using FeynCalc 
\cite{Mertig:1990an} and LoopTools \cite{Hahn:1998yk}. Then we take the experimentally 
measured values of these parameters \cite{Baak:2011ze} and perform a $\chi^2$-fit to six 
parameters: the four combinations of masses 
\begin{eqnarray}
m_q \equiv (m_{t'} + m_{b'})/2 \; , & \quad & 
\Delta_q \equiv m_{t'} - m_{b'} \; , \nonumber \\
m_\ell \equiv (m_{\nu'} + m_{\tau'})/2 \; , & \quad &
\Delta_\ell \equiv m_{\nu'} - m_{\tau'} \; , \nonumber 
\end{eqnarray}
the Higgs mass $m_h$, and the mixing $\sin \theta_{34}$. We take the ranges of $m_q$ and 
$m_\ell$ to be those allowed by the unitarity constraints and the direct search bounds stated 
above. For the other parameters, we take $|\Delta_q| < 200$ GeV, $|\Delta_\ell| < 200$ GeV, 
$100$ GeV $< m_h < 800$ GeV, $|\sin\theta_{34}| < 0.3$, unless explicitly specified otherwise. 

We present our results in terms of the goodness-of-fit contours for the joint estimates of 
two parameters at a time, where the other four parameters are chosen to get the minimum of $\chi ^2$. For the 
purposes of this investigation, we show contours of $p = 0.0455$, which correspond 
to $\chi^2 = 6.18$, or a confidence level(C.L) of 95\%. 
%
%
\section{Constraints on the mass splittings $\Delta_q$ and $\Delta_\ell$}
\label{sec:numerical}

We now focus on the constraints on the mass splittings $\Delta_q$ and $\Delta_\ell$. 
We scan over the allowed values of the other parameters, and take only those parameter sets 
that are consistent with all the data currently available. 

\subsection{Constraints on $\Delta_q$}
\label{sec:delta-q}

\begin{figure*}[t!]
\includegraphics[keepaspectratio=true,scale=0.4]{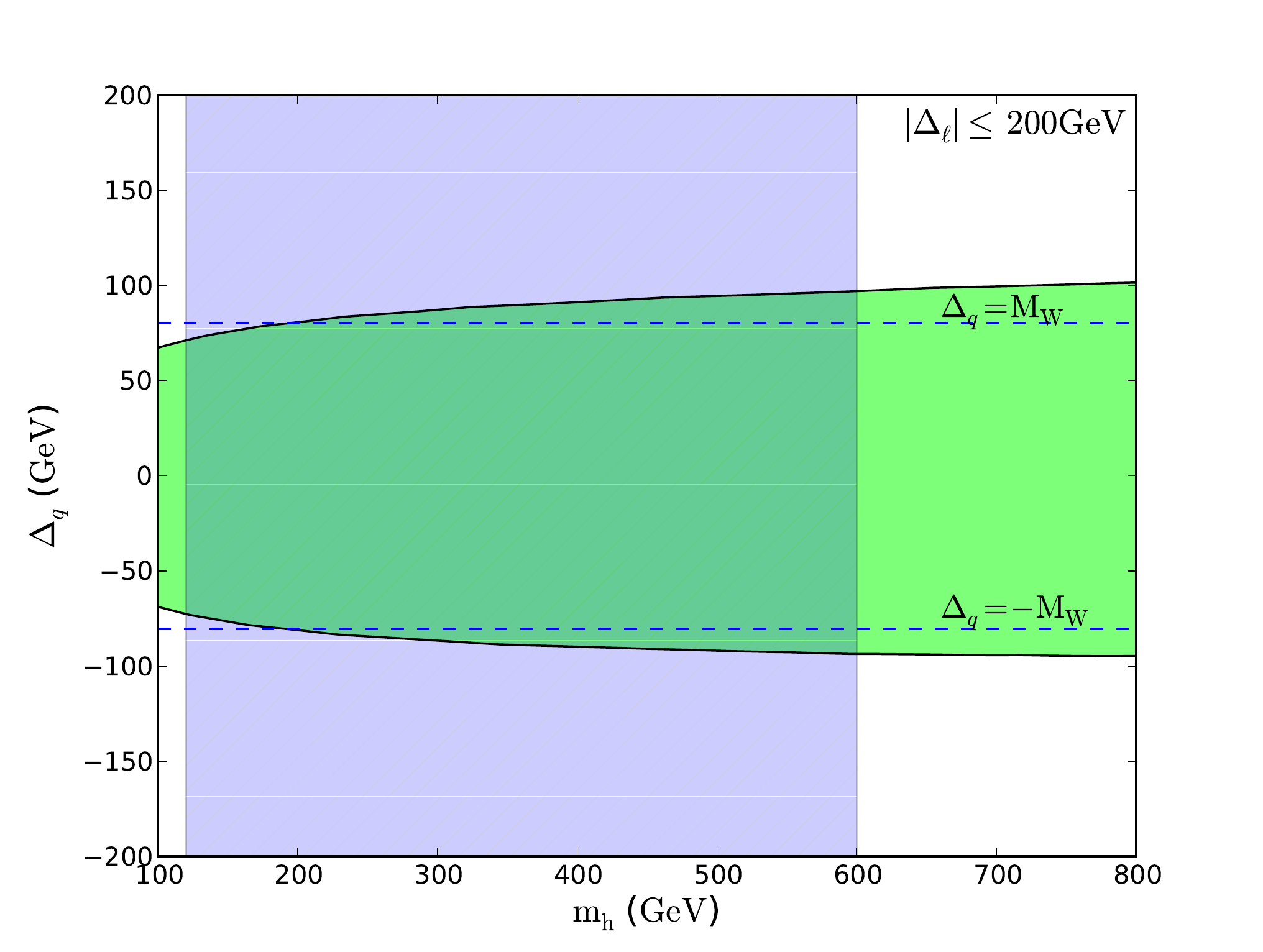}
\includegraphics[keepaspectratio=true,scale=0.4]{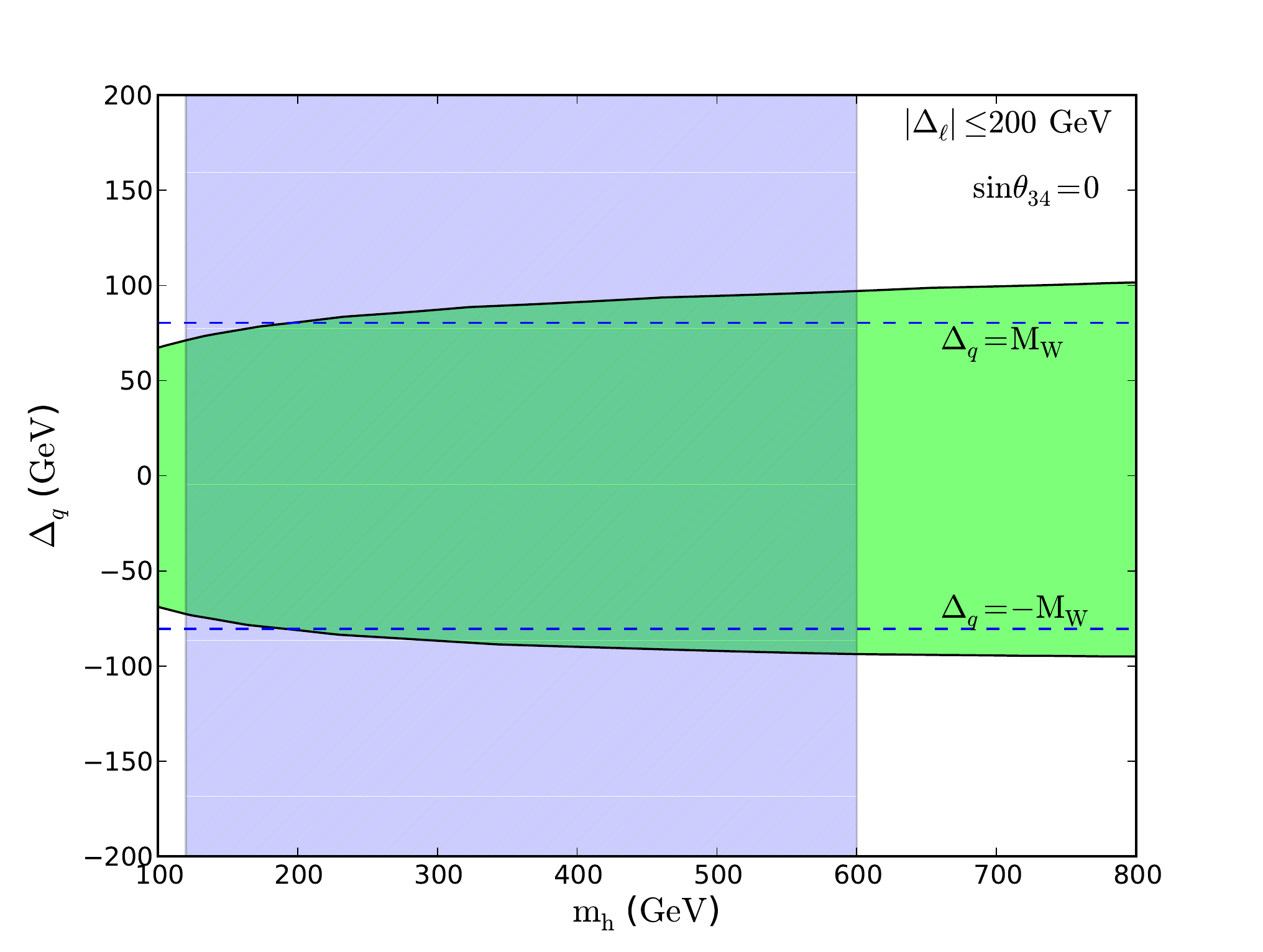}
\includegraphics[keepaspectratio=true,scale=0.4]{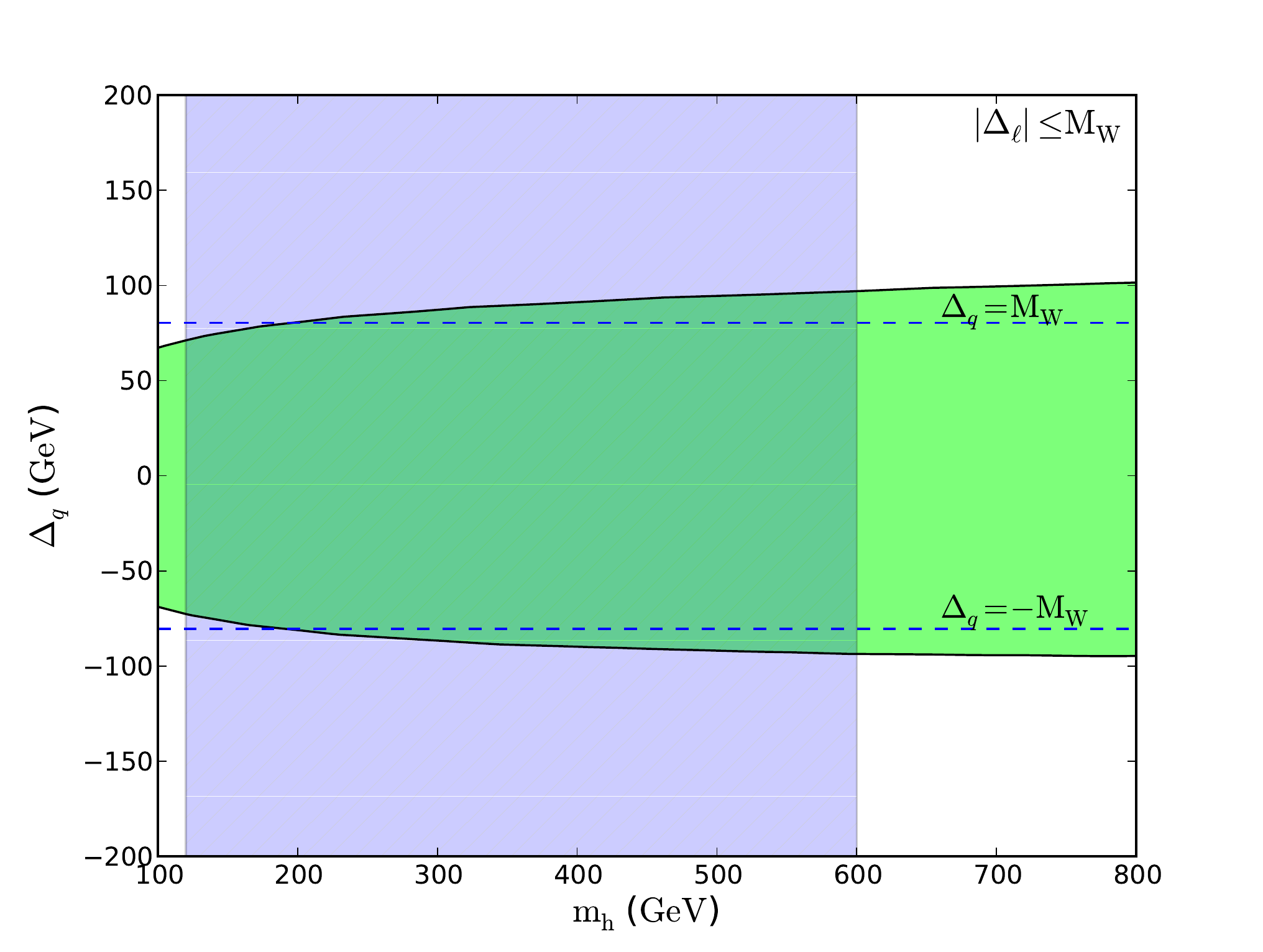}
\includegraphics[keepaspectratio=true,scale=0.4]{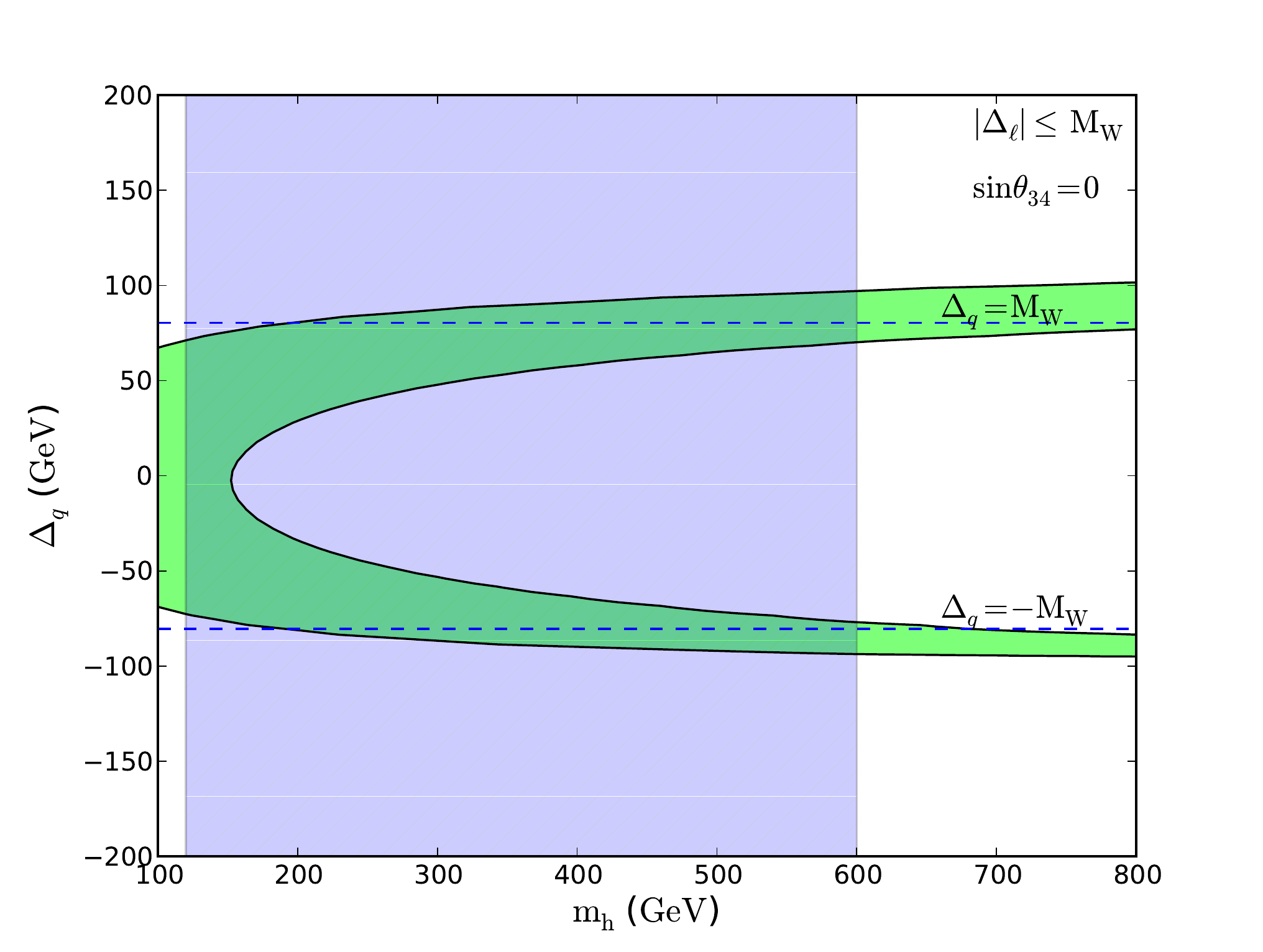}
\caption{The 95\% C.L. allowed regions in the $m_h$ -- $\Delta_q$ plane. 
In the top two panels, $\Delta_\ell$ is varied over $|\Delta_\ell| \leq 200$ GeV, while 
in the bottom two panels $\Delta_\ell$ is restricted to be less than $M_W$. In the left two 
panels, $\theta_{34}$ is varied over sin$\theta_{34} \leq 0.3$, while in the right 
two panels, $\theta_{34}$ has been fixed to zero (no mixing). The parameters $m_q$ and 
$m_l$ are varied over their complete allowed range. The grey shaded region is excluded 
at 95\% by the LHC data. The dashed blue lines correspond to $|\Delta_q| = m_W$.} 
\label{mu4-md4}
\end{figure*}

In the top left panel of Fig. \ref{mu4-md4}, we show the 95\% C.L. contours in the 
$m_h$ -- $\Delta_q$ plane marginalizing over other new-physics (NP) parameters 
($m_q, m_l, \Delta_\ell, \theta_{34}$). 
From this panel, it is observed that at large $m_h$ values, the value of 
$|\Delta_q|$ can exceed $M_W$. 

We now explore the effect of lepton mass splitting 
($\Delta_\ell$) and the fourth-generation quark mixing ($\theta_{34}$) in more detail. 
The bottom left panel shows the effect of restricting $|\Delta_\ell|$ to $M_W$, while 
the top right panel shows the effect of vanishing $\theta_{34}$. It is observed that 
there is no significant change in the allowed parameter space. 

However, when both the 
conditions of vanishing $\theta_{34}$ and  $|\Delta_\ell| < M_W$ are imposed, the 
character of the constraints changes dramatically, as can be seen from the bottom right 
panel. In this case, not only is $|\Delta_q| > M_W$ allowed at large $m_h$, one has to have 
$|\Delta_q| \geq M_W$ for $m_h \gtrsim 800$ GeV. At such large $m_h$ values, then, either 
$|\Delta_\ell| >m_W$ or $|\Delta_q|> m_W$. At least one of the decays via $W$ emission 
($t'\rightarrow b'W$, $b'\rightarrow t'W$, $\tau'\rightarrow \nu'W$ and 
$\nu'\rightarrow \tau 'W$) then must take place.

In most analyses of a fourth-generation scenario with a single higgs doublet, $|\Delta _q|$ 
had been assumed to be $\lesssim 75$ GeV due to the need to satisfy precision EW constraints. 
As seen above, the LHC exclusion of Higgs masses upto $m_h \sim 600$ GeV in fact allows 
larger mass differences in the fourth-generation quark doublet. Since $|\Delta _q|>M_W$ is 
allowed, the channel  $t'\rightarrow b'W$, or $b'\rightarrow t'W$ becomes allowed. 
This condition will have strong implications for the direct searches of the fourth 
generation scenario. 

%

The correlation between $\Delta_q$ and $\Delta_\ell$ observed above may be understood 
analytically as follows. First, note that the functions $T(x_u,x_d)$ in Eq.~(\ref{T-x1x2}) 
are positive semidefinite \cite{Peskin:1991sw}. The only negative contribution to $\Delta T_4$ 
of fermions is then from the first term of Eq.~(\ref{T:fermion}), however it is compensated 
by the next term which is larger in magnitude and positive. [In particular, when 
$x_{t'} \gg x_b$, the first two terms in $\Delta T_4$ add up to a positive quantity: 
$3 s^2_{34} (x_{t'}-x_{t}) /(16\pi s^2_W c^2_W)$.] 
Thus, the contribution from fermions to $\Delta T_4$ is positive for all values of fermion 
masses. On the other hand, Eq.~(\ref{T:higgs}) shows that the contribution from Higgs to 
$\Delta T$ is negative for $m_h > \tilde{m}_{h}=120$ GeV. In order to be consistent with 
precision electroweak data, the negative contribution to $\Delta T_H$ from Higgs should 
be compensated adequately by the positive contribution $\Delta T_4$ from fermions. 
(Although we also have $S, U$ parameters, the effect on $T$ dominates the behavior of our 
results.) This contribution comes from a combination of $T(x_u,x_d)$ in the quark and lepton 
sectors, leading to a strong correlation between the quark and lepton mass splittings.

When $\theta_{34}$ is zero or extremely small, the only contributions to $\Delta T_4$ 
are from $T(x_{t'},x_{b'})$ and $T(x_{\nu'}, x_{\tau'})$. These quantities then have 
to be sufficiently large to compensate for the large $\Delta T_H$ appearing at large $m_h$, 
necessitating a large mass splitting either in quark or in lepton sector. As the Higgs mass 
increases, the compensating contribution $\Delta T_4$, and hence the required mass 
splittings, also increase, exceeding $M_W$ for $m_h \gtrsim 800$ GeV. On the other hand, 
when $\theta_{34}$ is near its maximum allowed value of $\sin \theta_{34} = 0.3$, the first 
three terms in $\Delta T_4$ also contribute, as a result of which the mass splitting in 
fourth-generation quarks is restricted and cannot exceed $M_W$.

Based on the above discussion, the features of Fig.~\ref{mu4-md4} can be easily understood. 
The insensitivity of the allowed values of $|\Delta _q|$ to $|\sin\theta _{34}|$ in the top 
panels is mainly due to the fact that the lepton mass-splitting $|\Delta _{\ell}|$ is varied 
over a sufficiently large range. This ensures that the contribution of Higgs to $\Delta T$ 
is compensated by the contribution from fermions even in the absence of an enhancement of 
quark contribution by a nonzero value of $\sin\theta _{34}$. Now in the bottom-right panel, 
$|\Delta_{\ell}|$ is restricted to be less than $M_W$ along with $\sin\theta _{34}=0$. 
Here the absence of an enhanced quark contribution to $\Delta T_4$ due to 
$\sin\theta _{34}=0$, as well as insufficient contribution from leptons to $\Delta T_4$ due 
to the restriction on $|\Delta _{\ell}|$, lead to the exclusion of $|\Delta _q|\lesssim M_W$ 
at large $m_h$. In the bottom-left panel the excluded regions $|\Delta _q|\lesssim M_W$ become 
allowed as the contribution from the quarks to $\Delta T _4$ is enhanced by $\sin\theta _{34}$. 

Note that earlier works \cite{Kribs:2007nz,Chanowitz:2009mz,Eberhardt:2010bm} that had predicted the mass splitting in the quark sector to  be less than $M_W$  had focused on a light Higgs. That a lighter Higgs would only allow a small splitting should be clear from our figures and analytical arguments.
%
%
\subsection{Constraints on $\Delta_\ell$}
\label{sec:delta-ell}

\begin{figure*}[]
\includegraphics[keepaspectratio=true,scale=0.4]{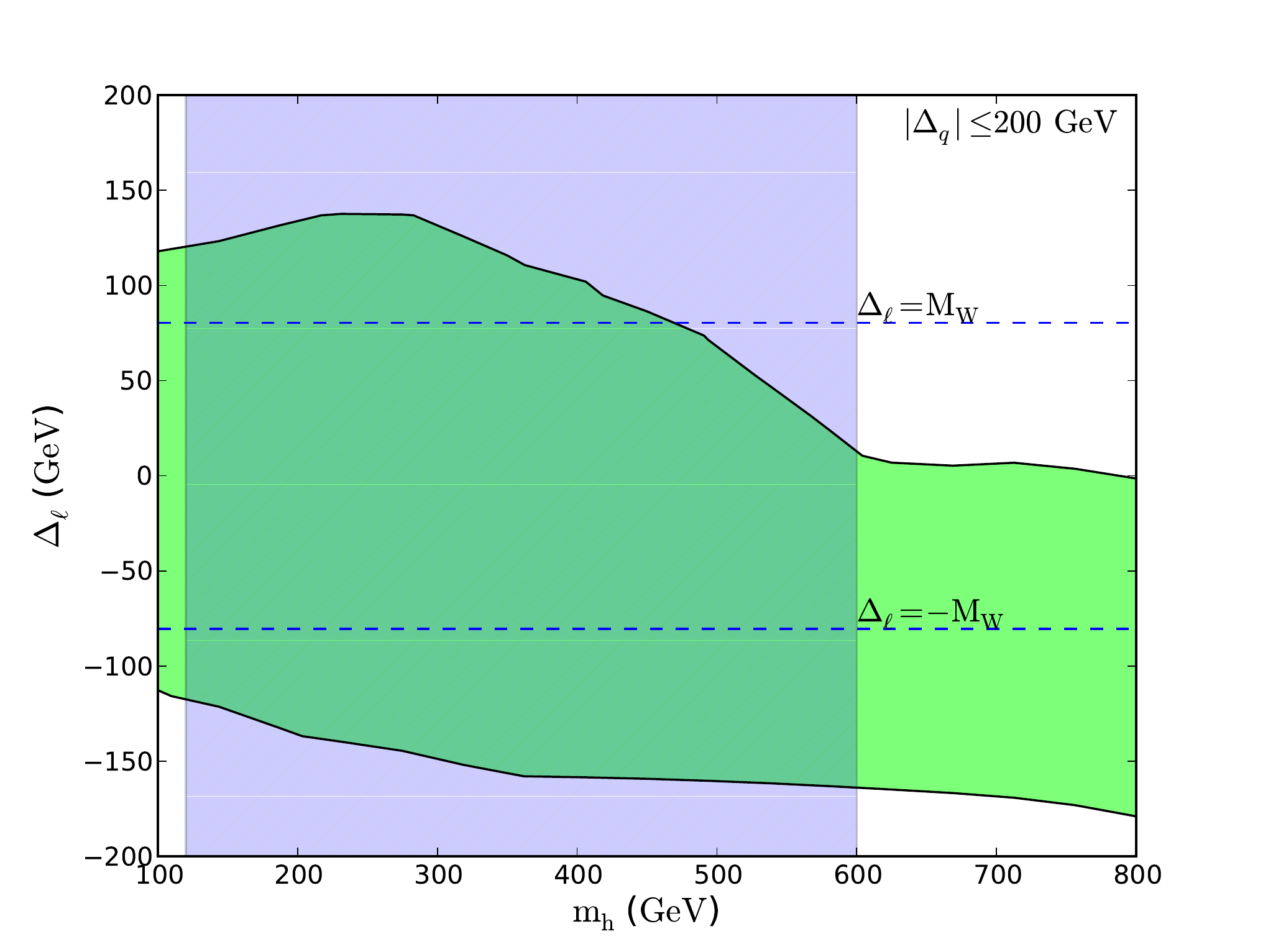}
\includegraphics[keepaspectratio=true,scale=0.4]{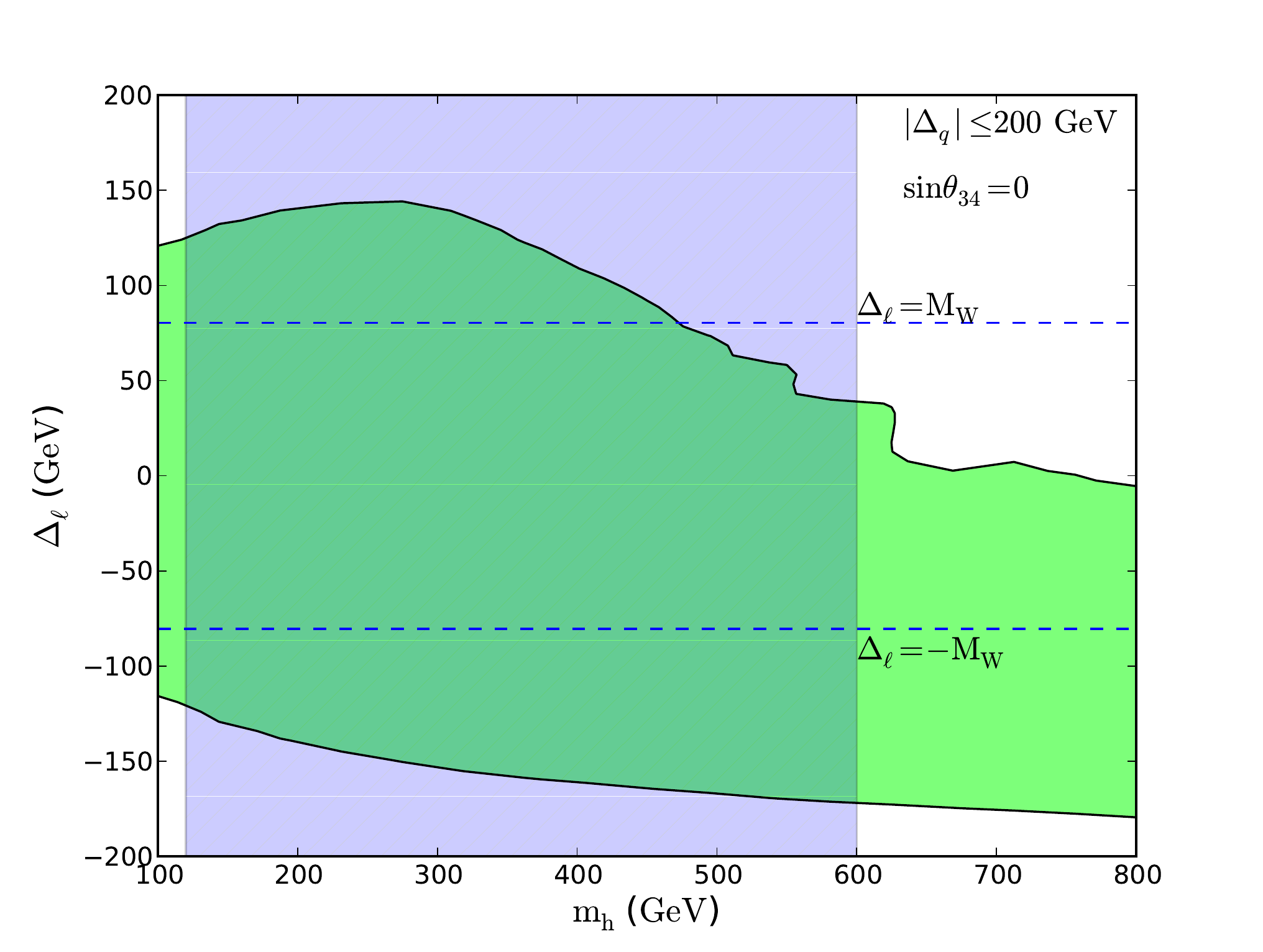}
\includegraphics[keepaspectratio=true,scale=0.4]{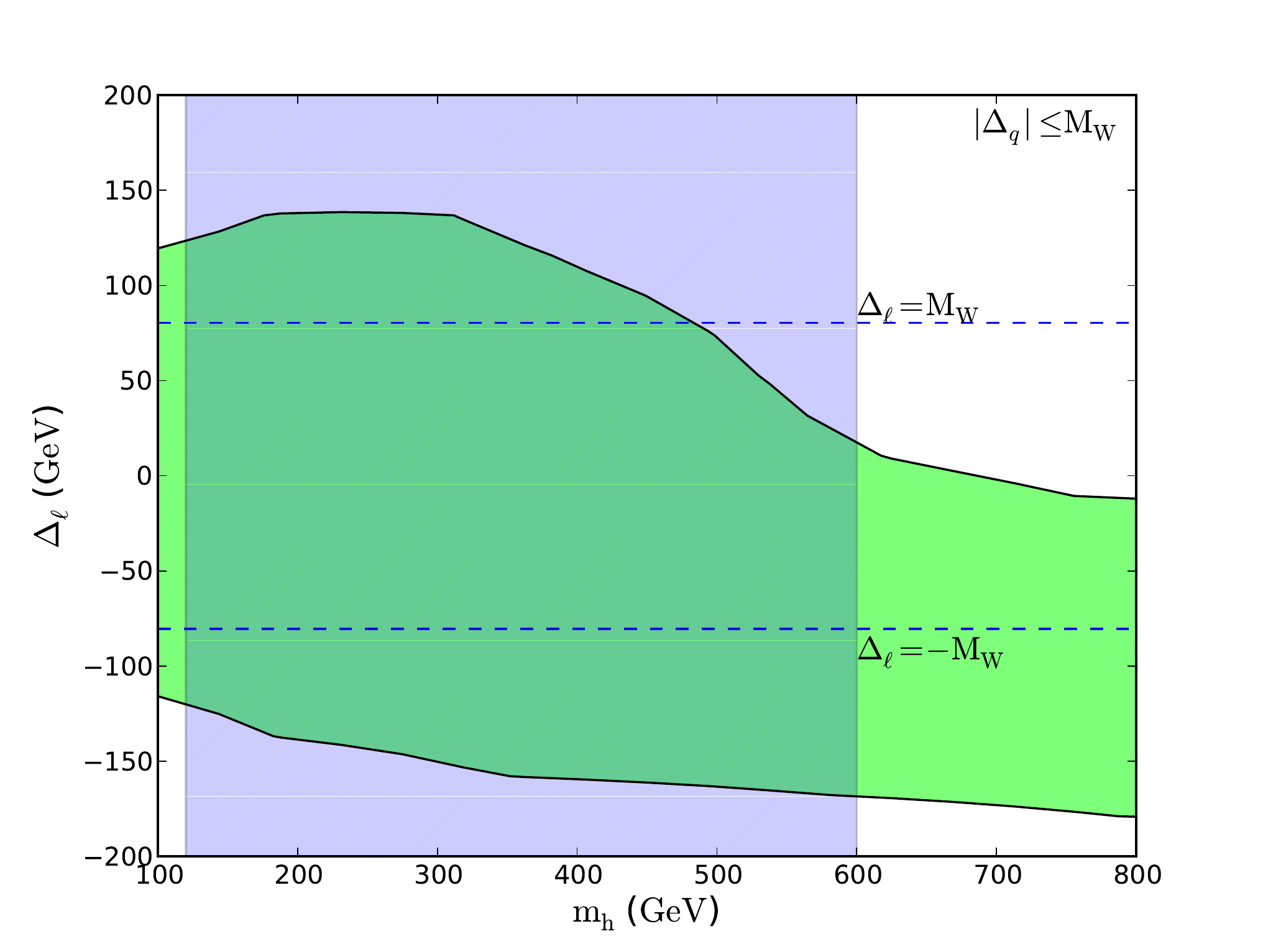}
\includegraphics[keepaspectratio=true,scale=0.4]{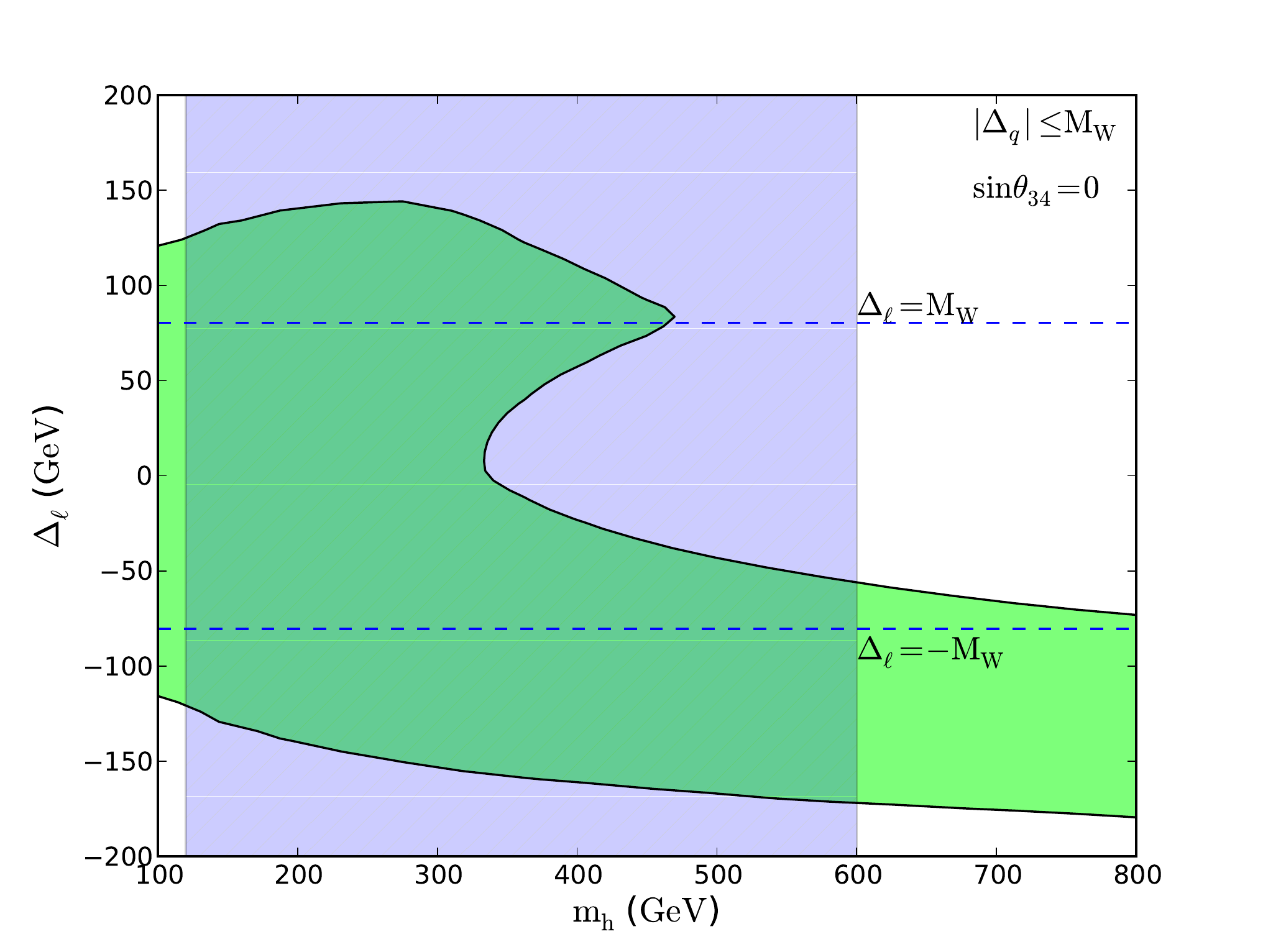}
\caption{The 95\% C.L. allowed regions in the $m_h$ -- $\Delta_\ell$ plane. 
In the top two panels, $\Delta_q$ is varied over $|\Delta_q| \leq 200$ GeV, while 
in the bottom two panels $\Delta_q$ is restricted to be less than $M_W$. In the left two 
panels, $\theta_{34}$ is varied over sin$\theta_{34} \leq 0.3$, while in the right 
two panels, $\theta_{34}$ has been fixed to zero (no mixing). The parameters $m_q$ and 
$m_l$ are varied over their complete allowed range. The grey shaded region is excluded 
at 95\% by the LHC data. The dashed blue lines correspond to $|\Delta_\ell| = m_W$.} 
\label{mnu4-me4}
\end{figure*}

In the top-left panel of Fig.~\ref{mnu4-me4}, we show the 95\% C.L. contours in the 
$m_h$ -- $\Delta_\ell$ plane, marginalizing over other NP parameters($m_q, m_l, \Delta_q, 
\theta_{34}$). It is observed that while at low $m_h$ values $\Delta_\ell$ can take 
any sign, at large $m_h$ values it is necessarily negative, i.e. $m_{\tau'} > m_{\nu'}$. 
Moreover, $|\Delta_\ell|$ can take values as large as 180 GeV. 

The effect of setting 
$\sin\theta _{34}=0$ on the allowed values of $|\Delta_\ell|$ is not significant, as 
can be seen in the top-right panel. If $|\Delta_q|$ is restricted to be less than $M_W$, 
while allowing any mixing sin$\theta_{34} \leq 0.3$, the parameter space is again not 
affected much as the bottom left panel shows. However, if $|\Delta_q|$ is restricted 
to be less than $M_W$, along with setting $\sin\theta _{34}=0$ as shown in the 
bottom-right panel, $|\Delta_\ell|$ is required to be large in magnitude. For $m_h \gtrsim 
800$ GeV, $|\Delta_\ell| > M_W$ and the decay $\tau' \to \nu' W$ is bound to occur. 
The features of the four panels regarding the role of $\sin\theta _{34}$ and $|\Delta_q|$ 
can be understood by the arguments given in the previous section.

%
%
\begin{figure*}[]
\includegraphics[keepaspectratio=true,scale=0.4]{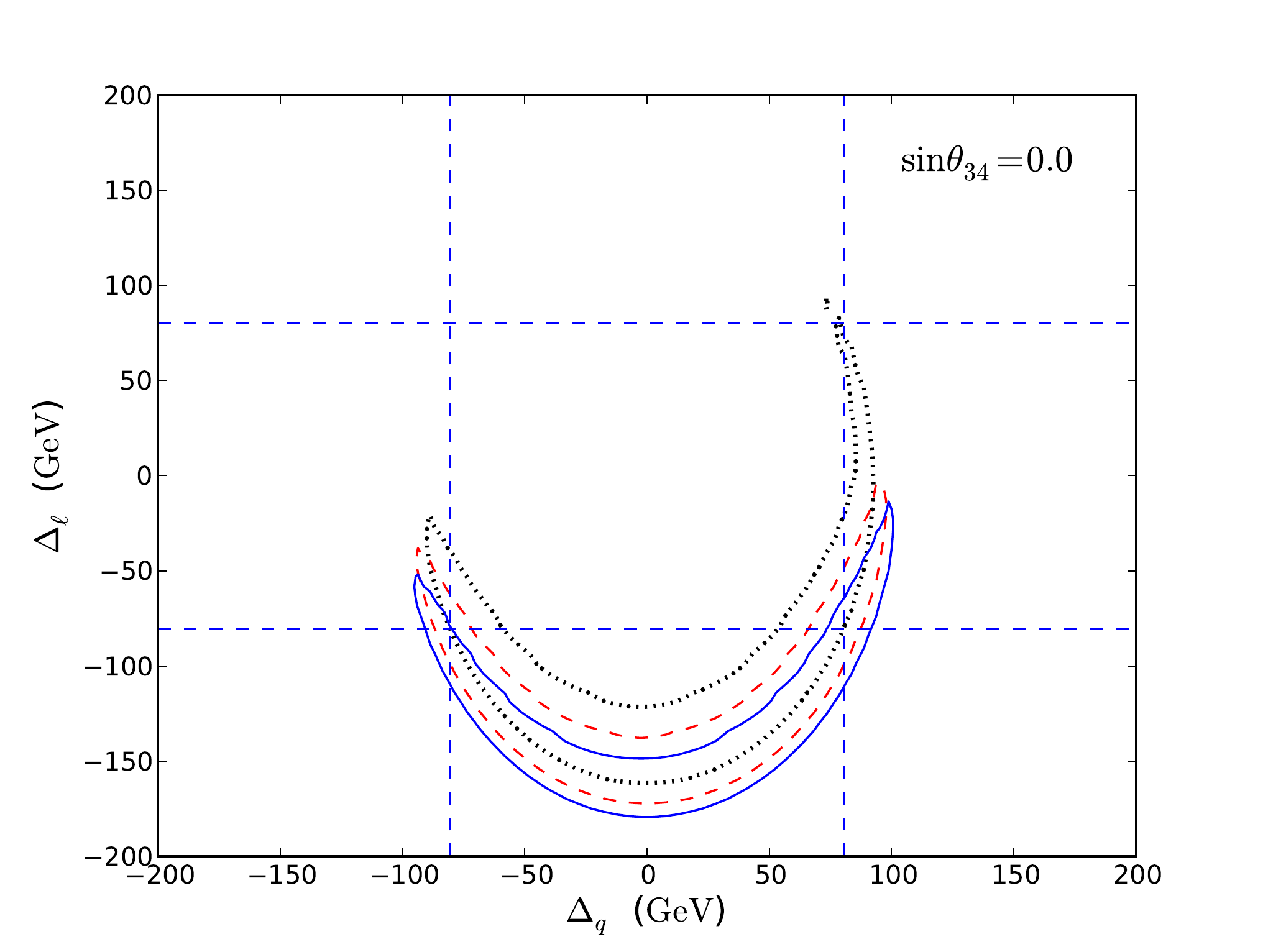}
\includegraphics[keepaspectratio=true,scale=0.4]{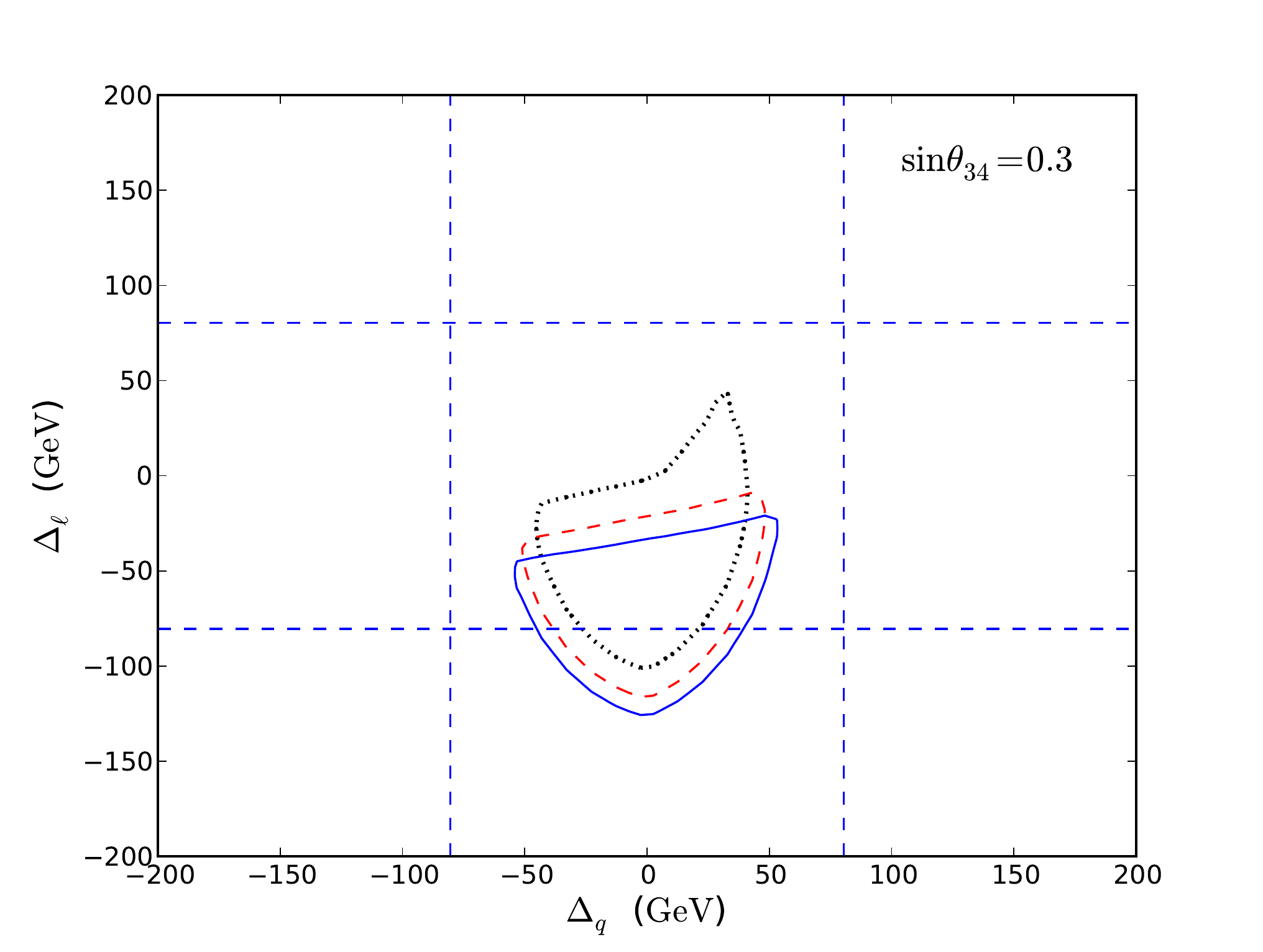}
\caption{The 95 \% C.L contours in the $\Delta_q$ -- $\Delta_\ell$ plane, for $m_h=400$ GeV 
(black dotted lines), $m_h=600$ GeV (red dashed lines) and $m_h=800$ GeV (blue/solid), respectively. 
The left (right) panel shows the results when $\sin \theta_{34}=0.0 (0.3)$. All the other 
parameters are varied over their 2$\sigma $ allowed ranges. The vertical blue dashed lines correspond to $|\Delta_q|= M_W$, 
while the horizontal green (dotted) lines correspond to $|\Delta_\ell|= M_W$.}
\label{Fig0:1}
\end{figure*}

The negative sign of $\Delta_\ell$ may be understood as follows. The Higgs contribution 
$\Delta S_H$ to the $S$ parameter is always positive, as can be seen from Eq.~(\ref{S-higgs}). 
It increases as $m_h$ increases. Since leptons have a negative hypercharge, the contribution 
to $S$ parameter from leptons can be reduced if $m_{\tau'}>m_{\nu'}$ for appropriate quark masses.

The interplay of the contributions to $\Delta S$ from the fourth-generation leptons and from 
the Higgs is also responsible for the asymmetry in the allowed region about 
$|\Delta _{\ell}|=0$ in the $\Delta _{\ell}$ -- $m_h$ plane. The $T$ parameter is approximately 
symmetric with respect to the masses of up and down-type fermions when the mass difference 
of fermions is small compared to their masses, as can be seen from Eq.~(\ref{Tdeg}). But for 
large $m_h$, minimizing the leptonic contributions to $S$ parameter becomes important for 
consistency. This causes the allowed regions to prefer $\Delta _{\ell}<0$ compared to 
$\Delta _{\ell}>0$, even though the $T$ parameter tends to produce symmetric allowed regions.

In contrast to leptons, for quarks, the $T$ parameter becomes important in constraining 
$|\Delta _{q}|$, as the hyperchage of quarks is positive. This makes allowed regions symmetric 
about $\Delta _{q}=0$.
%
\subsection{Constraints on ($\Delta_q,\Delta_\ell$) and the effect of $\theta_{34}$}

The left panel of Fig.~\ref{Fig0:1} shows the allowed parameter space in the 
$\Delta_q$ -- $\Delta_\ell$ plane for $\theta_{34}=0$, for different $m_h$ values. 
It can be easily seen that with increasing $m_h$, the allowed difference 
$m_{\tau'}-m_{\nu'}$ increases. This is consistent with the arguments in 
Sec.~\ref{sec:delta-ell} that used the contribution to $\Delta S$ from fermions and Higgs. 
Also, when the lepton splitting $|\Delta_\ell|$ is small, the quark splitting $|\Delta_q|$ 
has to be large to compensate for the Higgs contribution to $\Delta T$, as argued earlier 
in Sec.~\ref{sec:delta-q}. Indeed at large enough $m_h$ values, the allowed region is 
outside the central square and hence always corresponds to $|\Delta_q| > M_W$ or 
$|\Delta_\ell|>m_W$, implying that the $W$-emission channel is necessarily active. 
Therefore, in case further direct constraints increase the lower bound on $m_h$ to be 
$\gtrsim 900$ GeV, the $W$-emission signal is not observed in either quark or lepton 
channel, and $\theta_{34}$ is restricted by independent experiments to be very small (say $\sin \theta _{34}<0.05$), then the model with four generations can be ruled out at 95\% confidence level.

The scenario when the mixing angle $\theta_{34}$ is significant is shown in the right 
panel of Fig.~\ref{Fig0:1}, where the value of $\theta_{34}$ corresponds to the current 
upper bound on it. It can be seen that in such a case, $|\Delta_q|>M_W$ is forbidden, 
while $|\Delta_\ell|>m_W$ is allowed.

Our results are consistent with those obtained earlier in \cite{Hashimoto:2010at}. 
In addition, we have shown the pattern of allowed mass differences of fermions as a 
function of $m_h$ and quark mixing.
%
\section{Collider Implications}
\label{sec:collider}
%
%
\begin{figure*}
\includegraphics[keepaspectratio=true,scale=0.4]{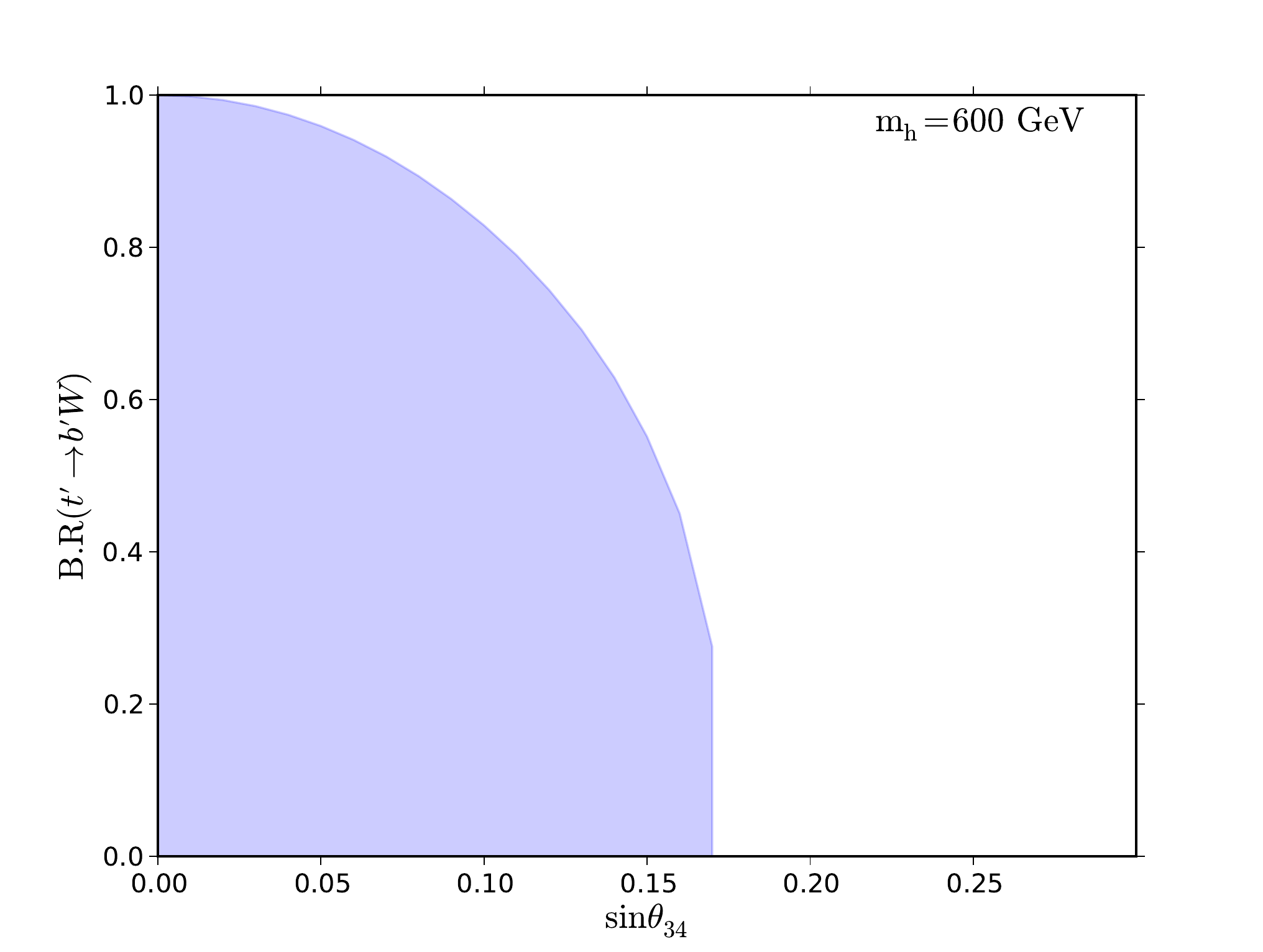}
\includegraphics[keepaspectratio=true,scale=0.4]{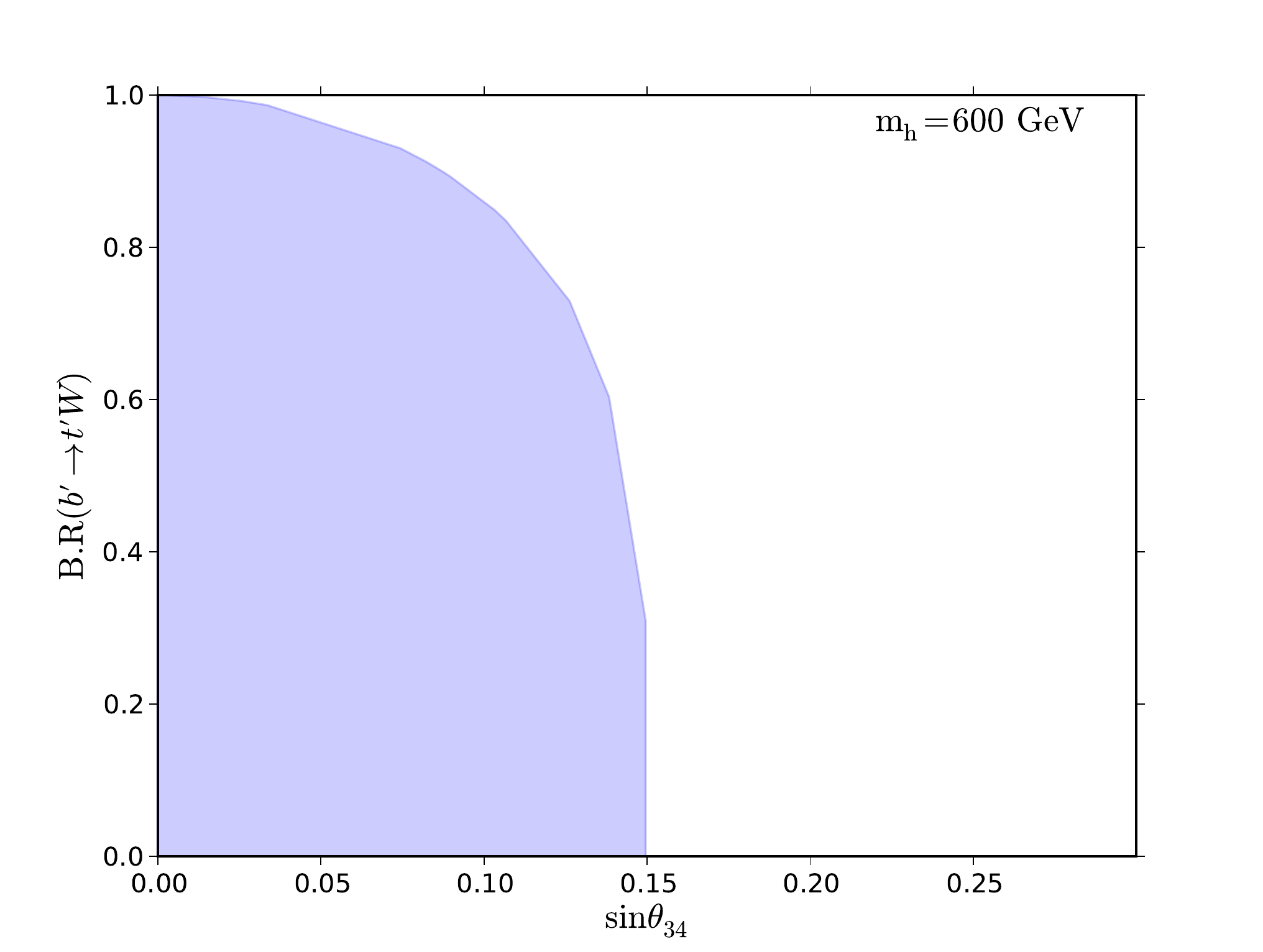}
\caption{Shaded region in the left (right) panel indicates the values of the 
branching ratios of $t' \to b' W$ ($b' \to t' W$) allowed at 95\% C.L. by all 
the constraints considered in this paper.} 
\label{BR}
\end{figure*}
%
\begin{figure}
\includegraphics[keepaspectratio=true,scale=0.4]{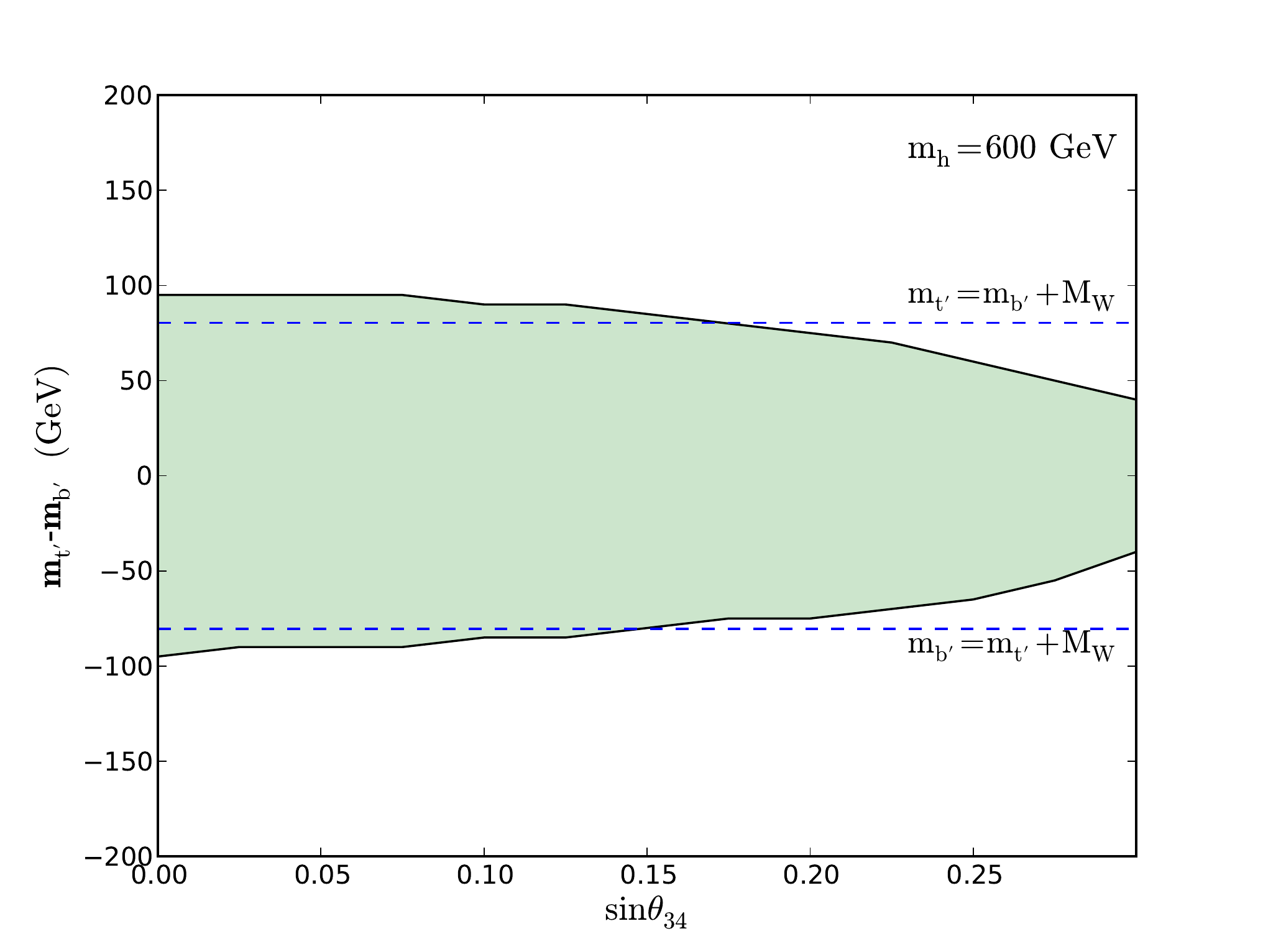}
\caption{Allowed values of $m_{t'}-m_{b'}$ (shaded region) for $m_{h}=600$ GeV as 
a function of $\sin \theta _{34}$ after marginalizing over the other NP parameters}
\label{dmqs34}
\end{figure}

According to our results, $m_{t'}>m_{b'}+M_W $ may be allowed for large $m_h$ values. 
In that case, the branching ratio BR($t'\to b'W$)  will depend on 
$\sin \theta _{34}$. Such a scenario was considered in \cite{Flacco:2011ym} to generalize 
the direct search experiment limits on $m_{t'}$ and $m_{b'}$ to include the effect of mixing 
of fourth-generation quarks to the three existing generations. In our case, we have zero 
mixing between fourth-generation quarks and the first two generations in contrast to 
the assumption of \cite{Flacco:2011ym}. But it does not affect the BR($t'\to b' W $) 
[although the decays $t'\to q W \ (q=d,s)$ will be forbidden] as long as the 
b-quark mass can be neglected. Therefore our use of the results of \cite{Flacco:2011ym} 
for the model-independent bounds on the quark masses is still justified. The consideration 
of $m_{t'}>m_{b'}+M_W$ scenario by \cite{Flacco:2011ym} was motivated by the result of 
\cite{Hashimoto:2010at} which stated that in a two-Higgs doublet model with a fourth 
generation, $m_{t'} - m_{b'}$ can be greater than $M_W$ and also be consistent with 
precision EW data. Reference \cite{Hashimoto:2010at} also shows that $m_{t'}-m_{b'}>M_W$ is 
possible with one Higgs doublet of SM4 when $m_{\tau'}-m_{\nu'}\leq M_W$. However, it 
assumes no 3-4 mixing in the quark sector. 
Similar conclusions hold also for the mass difference $m_{b'}-m_{t'}$, which may be 
greater than $M_W$, leading to the possibility of the decay channel $b' \to t' W$.
 
We have shown that $|m_{t'}-m_{b'}|>M_W$ is allowed even after marginalizing over the lepton 
masses and $\theta_{34}$. Our result, in addition to the result of \cite{Hashimoto:2010at}, 
justifies considering $|m_{t'}-m_{b'}|>M_W$ for interpreting direct search data on fourth 
generation quarks, as was done in \cite{Flacco:2011ym}. Our result also means that the 
conditions  $|m_{t'}-m_{b'}|>M_W$ can be met even in the case of one Higgs doublet. 
In Fig.\ref{BR}, we plot the allowed values of branching ratios of the decays 
$t'\to b' W $ and $b' \to t' W $ as functions of $\sin \theta _{34}$. 
One can easily see that  the branching ratio of the decay of a fourth-generation quark into 
another fourth-generation quark can be close 100\%. This emphasizes the need to consider 
these decay modes in direct search experiments which search for fourth-generation quarks.
 
Figure ~\ref{BR} shows that for $\sin \theta _{34} \gtrsim 0.15$ there exists no point 
in the parameter space which 
passes all the constraints (direct search, $S, T, U$) for which the decay 
($b'\to t' W $ or $t'\to b' W $) is possible. This can be understood from 
Fig.~\ref{dmqs34}, where we show the allowed values of $\Delta_q$ as a function of 
$\theta_{34}$. For large $\theta_{34}$ values, $\sin \theta_{34} \gtrsim 0.15$, the mass 
splitting goes below $M_W$ and the W-emission channel is forbidden.  
%
%
\section{Summary and Outlook}
\label{sec:concl}
%
We have explored the allowed mass spectra of fourth-generation fermions, calculating 
the constraints from direct searches at the colliders, the theoretical requirement of 
perturbative unitarity, and electroweak precision measurements. We take into account 
the masses of fourth-generation quarks as well as leptons, and possible mixing of the 
fourth-generation quarks with the third generation ones. The other mixings of the 
fourth-generation quarks and leptons are more tightly constrained, and hence neglected.

Our perturbative unitarity calculation with the inclusion of all the $J=0$ channels for 
$2 \to 2$ fermion scattering tightens the earlier upper bounds on fourth-generation quark 
masses by about 6\%, while keeping the constraints on the fourth-generation lepton masses 
unaffected. These bounds are relatively insensitive to the precision electroweak observables 
$S, T, U$. The mixing between the third and fourth-generation quarks is constrained 
primarily by the flavor-physics data, to $\sin \theta_{34} < 0.3$. The perturbative unitarity 
bounds depend only weakly on this mixing.

Performing a $\chi^2$-fit to the measured values of the precision electroweak parameters 
$S, T, U$, we find that large values of the Higgs mass, $m_h \gtrsim 600$ GeV as indicated 
by the current LHC data, allow the mass splitting between the fourth-generation quarks or 
leptons to be greater than $M_W$. In the case of the quark splitting $\Delta_q = m_{t'}-m_{b'}$, 
the possibility $|\Delta_q| > M_W$ starts being allowed at 95\% C.L. for $m_h \gtrsim 200$ GeV. 
For the lepton splitting $\Delta_\ell = m_{\nu'}-m_{\tau'}$, the possibility $\Delta_\ell > M_W$ 
is allowed at 95\% for $m_h \lesssim 450$ GeV, while $\Delta_\ell < - M_W$ is allowed for all 
values of $m_h$. Moreover, if $\theta_{34}$ is small, either $|\Delta_q| > M_W$ or 
$|\Delta_\ell|>M_W$ is necessary for values of $m_h$ as large as 800 GeV. We present 
correlations between the values of $\Delta_q$ and $\Delta_\ell$, as well as the constraints 
on them as functions of $m_h$ and $\theta_{34}$.

Most of the above observations may be explained qualitatively through the analytic expressions 
for the contribution to the $S$ and $T$ parameters by the Higgs and the fourth-generation fermions, 
and their interference. In particular, the requirement of $|\Delta_q| > M_W$ or $|\Delta_\ell|>M_W$ 
at large $m_h$ for small $\theta_{34}$, and the relaxation of this for large $\theta_{34}$, can 
be easily motivated. These expressions also allow an understanding of the asymmetric bounds 
on $\pm \Delta_\ell$, and why $m_{\tau'} > m_{\nu'}$ is necessary at large $m_h$. 
No such hierarchy of masses can be predicted in the quark sector.

The unique feature of our analysis are the simultaneous consideration of the lepton masses, 
the quark mixing, and the recent indication of the heavy Higgs. The major consequence of our 
result is the opening up of the $W$-emission channels $t' \to b' W$, $b' \to t'  W$, or 
$\tau' \to \nu' W$ for large values of $m_h$. This will have major implications for the direct 
collider searches for fourth-generation fermions which currently are performed assuming for example, $t'\rightarrow bW/b'\rightarrow tW $ as the dominant decay modes. Indeed, since the branching ratios of these decay 
modes, when kinematically allowed, are large, they can have impact on the currently stated exclusion bounds, which have been arrived at by assuming that these decays are not allowed. In order to get model-independent bounds on the masses of the fourth-generation fermions, it is necessary to analyze the data keeping open the possibility of large branching ratios  in the $W$-emission channels.

We also find in our analysis, that in case further direct constraints increase the lower bound on $m_h$ to be 
$\gtrsim 900$ GeV, the $W$-emission signal is not observed in either quark or lepton 
channel, and $\theta_{34}$ is restricted by independent experiments to be very small (say $\sin \theta _{34}<0.05$), then the model with four generations can be ruled out at 95\% confidence level.

In conclusion, the fourth generation is currently alive and well. However if the corresponding 
standard model Higgs is heavy, it presents the possibility of an early direct detection of 
the fourth-generation fermions and also affects the search strategies as well as possible exclusions of the 
fourth-generation scenario strongly. Either way will lead to an important step ahead in our understanding 
of the fundamental particles.

\begin{acknowledgements}
We would like to thank A.Lenz and A.Djouadi for useful comments. RMG  wishes to acknowledge the Department of Science and Technology of India,
for financial support  under the J.C. Bose Fellowship scheme under Grant No.
SR/S2/JCB-64/2007.
\end{acknowledgements}
%
%
%
%
%
%
%
%
\appendix
%
\section{Partial-wave amplitudes}
%

In this appendix, we give the expressions for the $J=0$ partial-wave amplitudes of 
$2\rightarrow 2$ scattering processes of the heavy quarks of the SM4. Unitarity of 
the S-matrix and the validity of perturbative expansion of the S-matrix at high 
center-of-mass energies constrain the behavior of scattering amplitudes at high 
center-of-mass energies. For example, in the case of $2\rightarrow 2$ scattering of 
scalars, the tree-level amplitude is restricted to less than unity, $|M_0|<1$. If 
the scattering particles have other quantum numbers, the tree-level amplitudes form a 
matrix in the  space of the quantum numbers. The analogous criterion for the 
perturbative unitarity of the S-matrix will be that the absolute value of the maximum 
eigenvalue of the matrix-valued amplitude should be less than unity. 

In the case of SM4, the  $J=0$ partial-wave amplitude  of $2\rightarrow 2$ fermion 
scattering receives contribution from processes of the type $F\bar{F}\rightarrow 
F' \bar{F}'$ where $F$ and $F'$ are two heavy fermions of the SM4. The 
scattering amplitudes depend on the helicity configurations of the initial and final 
state fermions. Since the couplings of fermions to bosons in the SM are of scalar, vector 
or axial-vector type, the number of helicity configurations at high center-of-mass 
energies which have nonvanishing $J=0$ amplitudes reduces to four: $\{++\rightarrow ++\}, 
\{++\rightarrow --\}, \{--\rightarrow ++\}, \{--\rightarrow --\}$. 
The $2\rightarrow 2$ scattering amplitude of quarks is a matrix with both helicity and color 
indices.  We consider only the amplitudes where the initial and final states are 
color-neutral and charge-neutral.

Let $t, t', b'$ be denoted by the indices $i=1,2,3$, respectively, and let $m_i$  be their masses.
In the limit where the center-of-mass energy of the scattering $s\gg m_im_j$, the tree-level 
amplitudes for the processes $F\bar{F}\rightarrow F'\bar{F}'$ may be written in the 
form of a $30 \times 30$ matrix $M$. This matrix may be conveniently represented in the basis 
\begin{eqnarray}
& & \left( 
t_{+} ^R \ \overline{t_{+} ^R} , \; t_{-}^R \ \overline{t_{-} ^R} ,\; 
t_{+} ^{' R} \ \overline{t_{+} ^{'R}} ,\;  t_{-}^{'R} \ \overline{t_{-} ^{'R}} , \; 
b_{+} ^{'R} \ \overline{b_{+} ^{'R}} ,\;  b_{-}^{'R} \ \overline{b_{-} ^{'R}} ,\right. \nonumber \\
& & \left. t_{+}^R\ \overline{t_{+}^{'R}}, \; t_{-}^R\ \overline{t_{-}^{'R}},\; t_{+}^{'R}\ \overline{t_{+}^{R}},\; t_{-}^{'R}\ \overline{t_{-}^{R}},\right. \nonumber \\
& & \left. t_{+} ^G \ \overline{t_{+} ^G} , \; t_{-}^G \ \overline{t_{-} ^G} , \; t_{+} ^{' G} \ \overline{t_{+} ^{'G}} ,\;  t_{-}^{'G} \ \overline{t_{-} ^{'G}} , \;
b_{+} ^{'G} \ \overline{b_{+} ^{'G}} ,\;  b_{-}^{'G} \ \overline{b_{-} ^{'G}}, \right. \nonumber \\
& & \left. t_{+}^G\ \overline{t_{+}^{'G}},\;t_{-}^G\ \overline{t_{-}^{'G}}, \; t_{+}^{'G}\ \overline{t_{+}^{G}},\; t_{-}^{'G}\ \overline{t_{-}^{G}},\right. \nonumber \\
& & \left. t_{+} ^B \ \overline{t_{+} ^B} , \; t_{-}^B \ \overline{t_{-} ^B} ,\; 
t_{+} ^{' B} \ \overline{t_{+} ^{'B}} ,\;  t_{-}^{'B} \ \overline{t_{-} ^{'B}} , \; b_{+} ^{'B} \ \overline{b_{+} ^{'B}} ,\;  b_{-}^{'B} \ \overline{b_{-} ^{'B}},\right. \nonumber \\
& & \left. t_{+}^B\ \overline{t_{+}^{'B}},\;t_{-}^B\ \overline{t_{-}^{'B}},\; t_{+}^{'B}\ \overline{t_{+}^{B}},\; t_{-}^{'B}\ \overline{t_{-}^{B}}  \right),
\nonumber
\end{eqnarray}
where $R, G, B$ represent the three colors.
In this basis, 
\begin{eqnarray}
M=
\begin{bmatrix}
A & B & B\\
B & A & B\\
B & B & A\\ 
\end{bmatrix}
\end{eqnarray}
\newline
Where $A,B$ are $10\times 10 $ matrices which describe the scattering amplitudes.
Taking the mixing between the third and fourth-generations into account,
the matrices $A$ and $B$ may be written as
\begin{widetext} 
\begin{eqnarray}
A =  -\frac{\sqrt{2}G_F}{8\pi} 
\begin{bmatrix}
m_{11} & 0 & m_{12} & 0 & 0 & -m_{13}x & 0 & 0 & 0 & 0 \\
0 & m_{11} & 0 & m_{12} & -m_{13}x & 0 & 0 & 0 & 0 & 0 \\
m_{21} & 0 & m_{22} & 0 & 0 & -m_{23}y & 0 & 0 & 0 & 0 \\
0 & m_{21} & 0 & m_{22} & -m_{23}y & 0 & 0 & 0 & 0 & 0 \\
0 & -m_{13}x & 0 & -m_{23}y & m_{33} & 0 & 0 & zm_{13} & 0 & z^{\ast}m_{23}\\
-m_{13}x & 0 & -m_{23}y & 0 & 0 & m_{33} & zm_{32} & 0 & z^{\ast}m_{31} & 0 \\
0 & 0 & 0 & 0 & 0 & z^{\ast}m_{32} & 0 & 0 & 0 & 0 \\
0 & 0 & 0 & 0 & z^{\ast}m_{31} & 0 & 0 & 0 & 0 & 0 \\
0 & 0 & 0 & 0 & 0 & zm_{31} & 0 & 0 & 0 & 0 \\
0 & 0 & 0 & 0 & zm_{32} & 0 & 0 & 0 & 0 & 0 \\
\end{bmatrix} , \nonumber \\
\end{eqnarray}

\begin{eqnarray}
B  =   -\frac{\sqrt{2}G_F}{8\pi}
\begin{bmatrix}
m_{11} & 0 & m_{12} & 0 & 0 & -m_{13} & 0 & 0 & 0 & 0 \\
0 & m_{11} & 0 & m_{12} & -m_{13} & 0 & 0 & 0 & 0 & 0 \\
m_{21} & 0 & m_{22} & 0 & 0 & -m_{23} & 0 & 0 & 0 & 0 \\
0 & m_{21} & 0 & m_{22} & -m_{23} & 0 & 0 & 0 & 0 & 0 \\
0 & -m_{13} & 0 & -m_{23} & m_{33} & 0 & 0 & 0 & 0 & 0 \\
-m_{13} & 0 & -m_{23} & 0 & 0 & m_{33} & 0 & 0 & 0 & 0 \\
0 & 0 & 0 & 0 & 0 & 0 & 0 & 0 & 0 & 0 \\
0 & 0 & 0 & 0 & 0 & 0 & 0 & 0 & 0 & 0 \\
0 & 0 & 0 & 0 & 0 & 0 & 0 & 0 & 0 & 0 \\
0 & 0 & 0 & 0 & 0 & 0 & 0 & 0 & 0 & 0 \\
\end{bmatrix} ,  \nonumber \\
\end{eqnarray}

where $m_{ij}=m_im_j$, $x=1-|V_{tb'}|^2$, $y=1-|V_{t'b'}|^2$, $z=V_{tb^{\prime}}V^{\ast}_{t^{\prime} b^{\prime}}$, and $z^{\ast}=V^{\ast}_{tb^{\prime}}V_{t^{\prime} b^{\prime}}$. Block-diagonalising M, we get 
\begin{eqnarray}
M'=\begin{bmatrix}
A+2B & \mathbf{0} & \mathbf{0}\\
\mathbf{0} & A-B & \mathbf{0}\\
\mathbf{0} & \mathbf{0} & A-B\\ 
\end{bmatrix}
\end{eqnarray}
The maximum eigenvalue is obtained from $A+2B$ \cite{Chanowitz:1978mv}, 
which is given by 
\begin{eqnarray}
A + 2 B  =  -\frac{3\sqrt{2}G_F}{8\pi} 
\begin{bmatrix}
m_{11} & 0 & m_{12} & 0 & 0 & -m_{13}\delta _x & 0 & 0 & 0 & 0 \\
0 & m_{11} & 0 & m_{12} & -m_{13}\delta _x & 0 & 0 & 0 & 0 & 0 \\
m_{21} & 0 & m_{22} & 0 & 0 & -m_{23}\delta _y & 0 & 0 & 0 & 0 \\
0 & m_{21} & 0 & m_{22} & -m_{23}\delta _y & 0 & 0 & 0 & 0 & 0 \\
0 & -m_{13}\delta _x & 0 & -m_{23}\delta _y & m_{33} & 0 & 0 & \delta _{z}m_{13} & 0 & \delta _z^{\ast}m_{23} \\
-m_{13}\delta _x & 0 & -m_{23}\delta _y & 0 & 0 & m_{33} & \delta _z m_{32} & 0 & \delta _z^{\ast}m_{31} & 0  \\
0 & 0 & 0 & 0 & 0 & \delta _z^{\ast}m_{32} & 0 & 0 & 0 & 0 \\
0 & 0 & 0 & 0 & \delta _z^{\ast}m_{31} & 0 & 0 & 0 & 0 & 0 \\
0 & 0 & 0 & 0 & 0 & \delta _zm_{31} & 0 & 0 & 0 & 0 \\
0 & 0 & 0 & 0 & \delta _zm_{32} & 0 & 0 & 0 & 0 & 0 \\
\end{bmatrix},  \nonumber \\
\end{eqnarray}
where $\delta _x=1-(1/3)|V_{tb'}|^2$,  $\delta _y=1-(1/3)|V_{t'b'}|^2$, $\delta _z=(1/3)V_{tb^{\prime}}V^{\ast}_{t^{\prime} b^{\prime}}$ and $\delta _z^{\ast}=(1/3)V_{tb^{\prime}}^{\ast}V_{t^{\prime} b^{\prime}}$.
\vspace{5mm}
\end{widetext}
Note that the presence of a nonzero mixing between the third and  fourth-generation quarks is responsible for the appearance of the channels  $t\bar{t'}\rightarrow b'\bar{b'}$ and $t'\bar{t}\rightarrow b'\bar{b'}$.These channels are directly proportional to the mixing matrix elements in contrast to other channels where the effect of mixing matrix elements is not significant. 

\bibliographystyle{apsrev}
\bibliography{sm4_references}

\begin{thebibliography}{82}
\bibitem[{\citenamefont{Rahatlou}(2012)}]{Rahatlou:2012hm}
\bibinfo{author}{\bibfnamefont{S.}~\bibnamefont{Rahatlou}}
  (\bibinfo{collaboration}{for the ATLAS Collaboration, for the CDF
  Collaboration, for the CMS Collaboration, for the D0 Collaboration})
  (\bibinfo{year}{2012}), \eprint{arXiv:1201.4810}.
\end{thebibliography}

\end{document}